\documentclass[sn-basic,iicol]{sn-jnl}

\graphicspath{ {images/} }

\usepackage{subcaption} 
\usepackage{multirow}

\jyear{2022}%

\begin{document}

\title[ ]{Remixing Functionally Graded Structures: Data-Driven Topology Optimization with Multiclass Shape Blending}

\author[1]{\fnm{Yu-Chin} \sur{Chan}}\email{ychan@u.northwestern.edu}
\author[1]{\fnm{Daicong} \sur{Da}}\email{dda@northwestern.edu}
\author[1,2]{\fnm{Liwei} \sur{Wang}}\email{iridescence@sjtu.edu.cn}
\author*[1]{\fnm{Wei} \sur{Chen}}\email{weichen@northwestern.edu}

\affil*[1]{\orgdiv{Dept. of Mechanical Engineering}, \orgname{Northwestern University}, \orgaddress{\city{Evanston}, \postcode{60208}, \state{IL}, \country{USA}}}
\affil[2]{\orgdiv{State Key Laboratory of Mechanical System and Vibration, School of Mechanical Engineering}, \orgname{Shanghai Jiao Tong University}, \orgaddress{\city{Shanghai}, \country{China}}}

\abstract{To create heterogeneous, multiscale structures with unprecedented functionalities, recent topology optimization approaches design either fully aperiodic systems or functionally graded structures, which compete in terms of design freedom and efficiency. We propose to inherit the advantages of both through a data-driven framework for multiclass functionally graded structures that mixes several families, i.e., classes, of microstructure topologies to create spatially-varying designs with guaranteed feasibility. The key is a new multiclass shape blending scheme that generates smoothly graded microstructures without requiring compatible classes or connectivity and feasibility constraints. Moreover, it transforms the microscale problem into an efficient, low-dimensional one without confining the design to predefined shapes. Compliance and shape matching examples using common truss geometries and diversity-based freeform topologies demonstrate the versatility of our framework, while studies on the effect of the number and diversity of classes illustrate the effectiveness. The generality of the proposed methods supports future extensions beyond the linear applications presented.} 


\maketitle

\section{Introduction}\label{sec:intro}
Multiscale mechanical structures present exciting functionalities and unprecedented performance ranging from global objectives like light-weighting, thermal conductivity and energy absorption~\citep{Da2019multiscaleBESO,jia2021crashworthiness,da2021data} to targeted local behaviors such as shape morphing or pattern reconfiguration for soft robots~\citep{Mirzaali2018softdevice,Boley2019shapeshift} and active airfoils~\citep{Lumpe2021reversemorph}. To design such complex structures, multiscale topology optimization (TO) has risen to prominence and flourished. 
While early research focused on periodic microstructures, two types of \textit{heterogenous} designs, in which neighboring microstructural topologies differ from each other, now surpass them in terms of performance: fully aperiodic systems and functionally graded structures (FGS). 

In this work, we aim to bridge the freedom of aperiodic designs with the efficiency and smooth interfaces of FGS. Their pros and cons are summarized in this section. Both typically follow a homogenization approach where the material properties of each element in the macroscopic structure are replaced by the effective properties of the microstructure at that location. This greatly expedites multiscale design as the performance can be evaluated at the macroscale, albeit with lowered accuracy since heterogeneous structures break the assumption of infinite periodicity~\citep{Andreassen2014hom2d}. The two approaches differ in that aperiodic systems can be composed of very different microstructures, whereas the change in the topologies and/or volume fractions of neighboring microstructures in FGS are deliberately designed to be continuous.

Although aperiodic systems allow immense design freedom, they come at the cost of explosive problem sizes. Two avenues of data-driven methods have emerged to counter this "curse of dimensionality". One assembles microstructures from pre-computed libraries via combinatorial optimization~\citep{Schumacher2015,Panetta2015}, and the other accelerates gradient-based design by creating deep learning (DL) models from massive datasets for dimension reduction and rapid property predictions~\citep{Wang2020cmame,Kumar2020spinodoid,xiao2021design}. 

These approaches are powerful, but require large overhead costs to build the datasets and models, and need careful strategies to select reasonably compatible neighboring microstructures. Even then, they may not achieve connectivity on par with FGS. Combined with the use of the effective properties, low connectivity can result in the manufactured performance deviating greatly from the optimized design. These drawbacks mean that, in their current state, data-driven methods are difficult to scale to large systems with complex physics, e.g., nonlinear mechanics. Compared to FGS, such approaches are less suitable when stress concentrations or expensive property simulations must be avoided. 

In contrast, some works have indicated that the continuous interfaces and more gradual topological change in FGS may be able to mitigate the errors from homogenization~\citep{Schumacher2015,Garner2019compatibility,Panesar2018graded}.
Functional grading can be further categorized into three camps: continuously varying volume fraction~\citep{Wang2018gradedmorph,Li2019tpms,Jansen2020hybridfgs}, topology~\citep{Kumar2020spinodoid,Sanders2021multilattice}, or a hybrid of both~\citep{Wang2020lvgp,Luo2021fixinterclass,xiao2021design}. Recent research is shifting towards the latter two, which have demonstrated that expanding the design space to include multiple topology types can considerably improve the structural performance. Our work belongs to the hybrid one, and we define a \textit{microstructure class} as a family of microstructures that possess the same overall topological concept but vary individually by volume fraction. Hence, the works in the last two categories can be termed \textit{multiclass}.

Within existing multiclass methods, the prevailing strategy is to treat FGS design as a multi-material TO problem by allocating each class to its own region with distinct boundaries. This assumes that the interfaces between classes are perfectly connected. As a result, most approaches pre-define a few mutually compatible classes~\citep{Wang2020lvgp,Luo2021fixinterclass} or fix their connections~\citep{Chu2019fixconn,Zhang2018krigingFGScomp}, which reduces computational cost and complexity but can yield suboptimal solutions.

Similar methods in the general multiscale TO field accomplish connected heterogeneous designs without the above simplifications by: (1) sharing design variables at interfaces~\citep{Liu2019subdomain}, (2) adding connectivity constraints~\citep{Du2018connectivity,Garner2019compatibility}, (3) controlling the change in the properties of intermediate microstructures~\citep{Zhou2019levelsetmorph}, (4) creating geometric gradations during pre- or post-processing~\citep{Sanders2021multilattice,Zhou2019levelsetmorph,Zobaer2020voidmorph}, and (5) interpolating random field representations of microstructures~\citep{Kumar2020spinodoid}. Of these, \cite{Liu2019subdomain}, \cite{Garner2019compatibility}, and \cite{Sanders2021multilattice} concurrently design the macrostructure as well as the distributions of multiple microstructures, and only \cite{Luo2021fixinterclass} also optimized the graded volume fractions. 
Moreover, many do not scale well with the number of classes. 

Merging the advantages of the two heterogeneous approaches, we propose a general TO framework for multiclass FGS that achieves smooth transitions between multiple microstructure classes without additional constraints, even if those classes are not compatible initially. The cornerstone of our approach is a novel multiclass shape blending scheme that generates new microstructures from a small set of predefined basis classes while guaranteeing feasibility. We define a design as \textit{feasible} when the microstructures are not only self-connected (i.e., have no disconnected features) but also well-connected to their neighbors. In addition, we desire the microstructures to meet a minimum feature size, as required in some manufacturing techniques.
Our framework departs from existing heterogeneous design methods in several ways: 
\begin{itemize}
    \item We create a continuous and low-dimensional microstructure representation by using the parameters of our shape blending scheme as design variables. This effectively transforms the microscale problem into a parametric one. While we do require predefined basis classes, this by no means restricts our design, as we allow the regions for each class and the boundaries between them to be blurred so that the microstructures anywhere in the FGS can be novel, i.e., not found in the initial basis classes.
    \item Our blending scheme integrates naturally into existing TO methods. In this work, we incorporate it with discrete and gradient-based TO, along with a new penalty that promotes diverse designs. These significantly reduce the cost of concurrent design with any number of microscale classes. Thus, through blending, our framework features design freedom near that of aperiodic methods while inheriting the efficiency of functionally graded design.
    \item The blending parameters serve as effective inputs for neural networks that predict the effective properties of new microstructures and can be re-used in multiple applications, further accelerating design. 
    \item Certain feasibility metrics (e.g., self-connectedness and minimum feature size) can be built into the blending scheme so that they do not need to be included explicitly in the TO problem. Furthermore, the flexibility of using basis classes permits designers to incorporate expert knowledge and eliminates the frustration of choosing compatible classes. If desired, manual selection can be removed altogether by extracting classes from open-source databases, e.g., using diversity metrics~\citep{Chan2020metaset}.
\end{itemize}

With compliance and shape morphing examples, we demonstrate the efficiency and inherent ability of our approach in designing multiscale structures with continuous transitions. The benefits of multiclass FGS are verified by utilizing both common truss-type and diverse freeform (topology-optimized) basis classes, and by comparing our results to designs in literature.

\section{Methodology}\label{sec:methods}
In this section, we introduce three crucial components in our approach: (1) multiclass shape blending and global interpolation schemes (Sec.~\ref{sec:method_blending}), (2) neural networks for property prediction (Sec.~\ref{sec:method_nn}), and (3) concurrent multiclass data-driven topology optimization (TO), which ties all of the methods into one framework (Sec.~\ref{sec:method_design}).

\subsection{Multiclass Shape Blending and Smooth Interpolation}\label{sec:method_blending}
\subsubsection{Background}\label{sec:method_blend_bg}
Our proposed shape blending scheme is heavily inspired by the computer graphics field, where morphing one geometric model into another has long been studied and utilized in, e.g., animation films and video games~\citep{sanchez2016thesis,Rohra2019imagemorph}. These methods have also supported applications like medical imaging~\citep{Carballido2005mrimorph} and metal-forming simulations~\citep{Thomas2020metalmorph}. In fact, evolving geometries through surface representations is the foundation of the level-set TO method~\citep{vanDijk2013lsto}, while combining shapes using distance fields is the bedrock of TO algorithms like Movable Morphing Components~\citep{Zhang2015newMMC}. 

Most closely related to the interest of this paper are blending techniques that use function representations (FReps) of shapes, which parametrize any geometric model as a series of operations (e.g., unions, differences, and intersections) performed on a set of primitives or basis geometries~\citep{sanchez2016thesis}. It is extremely flexible as it allows the bases to be defined by any representation, e.g., meshes or voxels, and any resolution. In our case, we represent our basis microstructure classes as continuous signed distance fields (SDFs), which are implicit function representations similar to level-sets. That is, whether the material is solid or void at any arbitrary point $(x,y)$ within a microstructure, $\boldsymbol{\Phi}(x,y,t)$, is determined by whether the SDF value is positive or negative, respectively.
The isovalue $t$ controls the isocontour of the field and therefore enables us to tune volume fractions. With this powerful representation, we can not only generate an entire family of microstructures over a continuous range of volume fractions, but also combine multiple SDFs to create novel classes.

Blending operations to mix SDFs have been studied for decades, beginning with the simple set-theoretic operation for the union of the function representations of two shapes, $\boldsymbol{\Phi}_1$ and $\boldsymbol{\Phi}_2$~\citep{Ricci1973settheoretic}: $\boldsymbol{\Phi}_{union}=\max(\boldsymbol{\Phi}_1,\boldsymbol{\Phi}_2)$.
Since then, many works have improved shape metamorphosis~\citep{sanchez2016thesis,Eisenberger2019shapecorresp,oring2020latentinterp}, but require intensive user-interaction and computational costs, and are, therefore, intractable or unsuitable for multiscale design.

\subsubsection{Multiclass Shape Blending Scheme}\label{sec:method_blend_blend}

\begin{figure*}[t]
    \centering
    \includegraphics[width=0.67\textwidth]{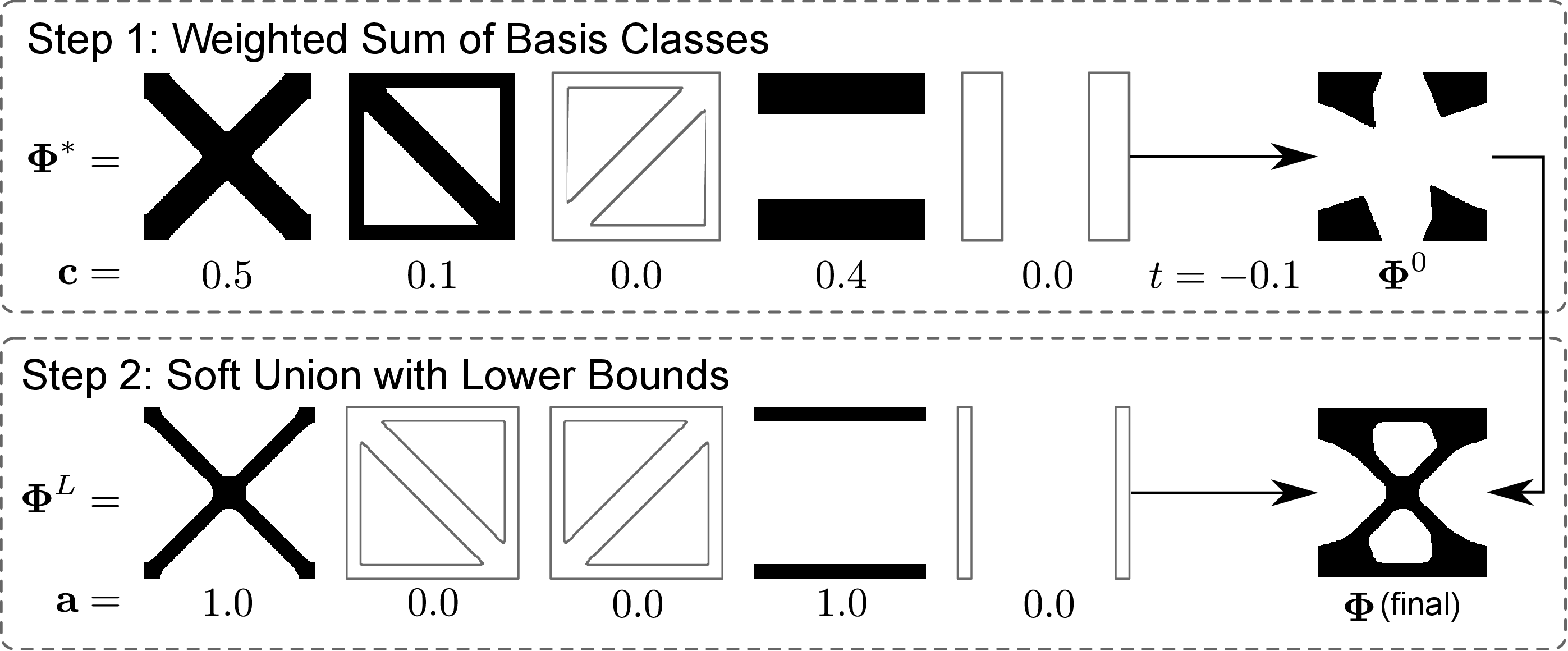}
    \caption{ Illustration of using multiclass shape blending scheme to generate a microstructure from truss basis classes. Outlined shapes represent classes whose weights are zero. Top row: Step 1 (Eq.~\ref{eq:blend_inner}), bottom row: Step 2 (Eq.~\ref{eq:blend_outer}). }
    \label{fig:blend_demo_unit}
\end{figure*}

Our proposed multiclass shape blending scheme is a combination of two simple techniques: (1) a weighted sum of basis classes based on cross dissolving, and (2) an activated union with the lower feasible bounds of each basis. The entire scheme is differentiable and efficient, which fits in well with gradient-based optimization algorithms. 

Crucial to our scheme are basis microstructure classes, which do not need to be mutually compatible. We denote the SDF representations of arbitrary basis classes as $\boldsymbol{\Phi}_d^B$ for $d=[1,D]$, where the shapes $\boldsymbol{\Phi}_d^B$ and total number of classes $D$ are defined by the user. Note that while we use truss-type bases in Figs.~\ref{fig:blend_demo_unit} and~\ref{fig:blend_demo} as demonstration, our method is applicable to any set of classes.

Prior to blending, we normalize the $D$ basis classes so that they can be mixed fairly and also find their lower feasible bounds. The following pre-processing steps only need to be performed once per set of bases:
\begin{enumerate} 
    \item Choose a common volume fraction, $v^*$, and find the representative SDF for each basis, $\boldsymbol{\Phi}_d^{*}=\boldsymbol{\Phi}_d^B+t_d^*$, such that it has $v^*$. We use the well-known bisection algorithm.
    \item Find the SDF of the lower feasible bound of each basis, $\boldsymbol{\Phi}_d^{L} = \boldsymbol{\Phi}_d^B+t_d^{L}$, using any desired feasibility metric. In this work, we set a minimum feature size of $4$ pixels for a $50 \times 50$ microstructure.
\end{enumerate}

After this one-time process, we can use $\boldsymbol{\Phi}_d^{*}$ and $\boldsymbol{\Phi}_d^{L}$ repeatedly for our proposed multiclass blending. The first step of the scheme is based on cross dissolving~\citep{Rohra2019imagemorph}, which can be simply defined as a linear interpolation between the source and target geometries. 
This can induce a double exposure effect where traces of both shapes co-exist in the blended result. 
While that causes unnatural morphing of, e.g., human faces, it organically achieves connected transitions between neighboring microstructures. 

We express multiclass interpolation as a weighted sum of the bases:
\begin{equation}\label{eq:blend_inner}
    \boldsymbol{\Phi}^0 = \sum_d^D c_{d} \boldsymbol{\Phi}_d^{*} + t,
\end{equation}
where $c_{d} \in [0,1]$. 
Although it effectively creates new classes of microstructures, this interpolation is agnostic to important geometrical features and can lead to broken shapes that have disconnected or thin features (see the top row of Fig.~\ref{fig:blend_demo_unit}). 

Therefore, to guarantee that the blended microstructures are sufficiently connected and feasible, we propose an additional step that enforces a lower feasible bound on blending. It also acts as an implicit constraint for simple manufacturing considerations, e.g., minimum feature sizes or volume fractions. 
This second step is an activated soft-max function, a continuous and differentiable extension of Ricci's set-theoretic union above:
\begin{equation}\label{eq:blend_outer}
    \boldsymbol{\Phi} = \frac{1}{\beta_2} \log{ \Big[ \exp{\big( \beta_2 \boldsymbol{\Phi}^0\big)} + \sum_d^D a_d \exp{\big( \beta_2 \boldsymbol{\Phi}_d^{L} \big)} \Big] },
\end{equation}
where $a_d = H(c_d)$ are the activated weight parameters using the Heaviside function:
\begin{equation}\label{eq:heav}
    H(c_d) = \frac{ \tanh{(\beta_2\eta_2)} + \tanh{(\beta_2(c_d-\eta_2))} }{ \tanh{(\beta_2\eta_2)} + \tanh{(\beta_2(1-\eta_2))} }.
\end{equation}
By setting the threshold $\eta_2>0$, the activation guarantees that at least one~--~but not all, or else low volume fractions would be difficult to attain~--~of the bases are feasible in each blended result. That is, in this step, only weights that are greater than $\eta_2$ are activated to equal one (shaded shapes in the second panel of Fig.~\ref{fig:blend_demo_unit}) while others are suppressed to zero (outlined shapes). We find that setting $\eta_2$ to the $75th$-percentile of the weights, $\mathbf{c}$, works quite well in promoting connected transitions.

With the two-step shape blending scheme, the representation of all possible blended microstructures, including those in the original basis classes, can be compactly expressed as the weight parameters. We can, therefore, formulate the microscale design variables as $\mathbf{c}^{(m)}$ for $m\in[1,M]$ desired optimal classes (set by the user), transforming the typically high-dimensional optimization problem into a simple and efficient parametric one that can still generate a wide range of microstructures. 

A final note regarding microstructure design is that, in practice, Eq.~\ref{eq:blend_inner} allows each $c_d^{(m)}\in[0,1]$ and can lead to $\sum_{d=1}^D c_d^{(m)} = 0$, which results in completely solid microstructures and occasional numerical issues. 
Moreover, it causes redundancy in the design space since taking $\boldsymbol{\Phi}^0 \geq 0$ to obtain the solid topology cancels out the least common denominator of the weights. There are numerous ways to enforce the sum of $c_d^{(m)}$ to be equal to one. We extend a multi-material interpolation schemes from our previous work~\citep{Chan2019multimat} to normalize the basis class weights as follows:
\begin{equation}\label{eq:class_interp}
    \tilde{\mathbf{c}}^{(m)} = \mathbf{z}^{(1)} + \sum^{D-1}_{j=1} \bigg[ \big(\mathbf{z}^{(j+1)} - \mathbf{z}^{(j)}\big) \prod^{j}_{k=1} c_k^{(m)} \bigg],
\end{equation}
where $\mathbf{z}^{(j)}$ are constant one-hot encoded vectors for each basis such that $z_i^{(j)}$ equals $1$ for $i=j$ and $0$ for all $i\neq j$. Subsequently, $c_d$ in Eqs.~\ref{eq:blend_inner} and~\ref{eq:heav} are replaced with $ \tilde{c}_d$. 
While our shape blending and design methods work well without Eq.~\ref{eq:class_interp}, for increased stability in the optimization process, and the added bonus of reducing the microscale design variables, $\mathbf{c}^{(m)}$, to size $[1 \times D-1]$, we use it in the remaining discussions.

\subsubsection{Integration with Multiscale Design}\label{sec:method_blend_global}

\begin{figure*}[th]
    \centering
    \includegraphics[width=0.78\textwidth]{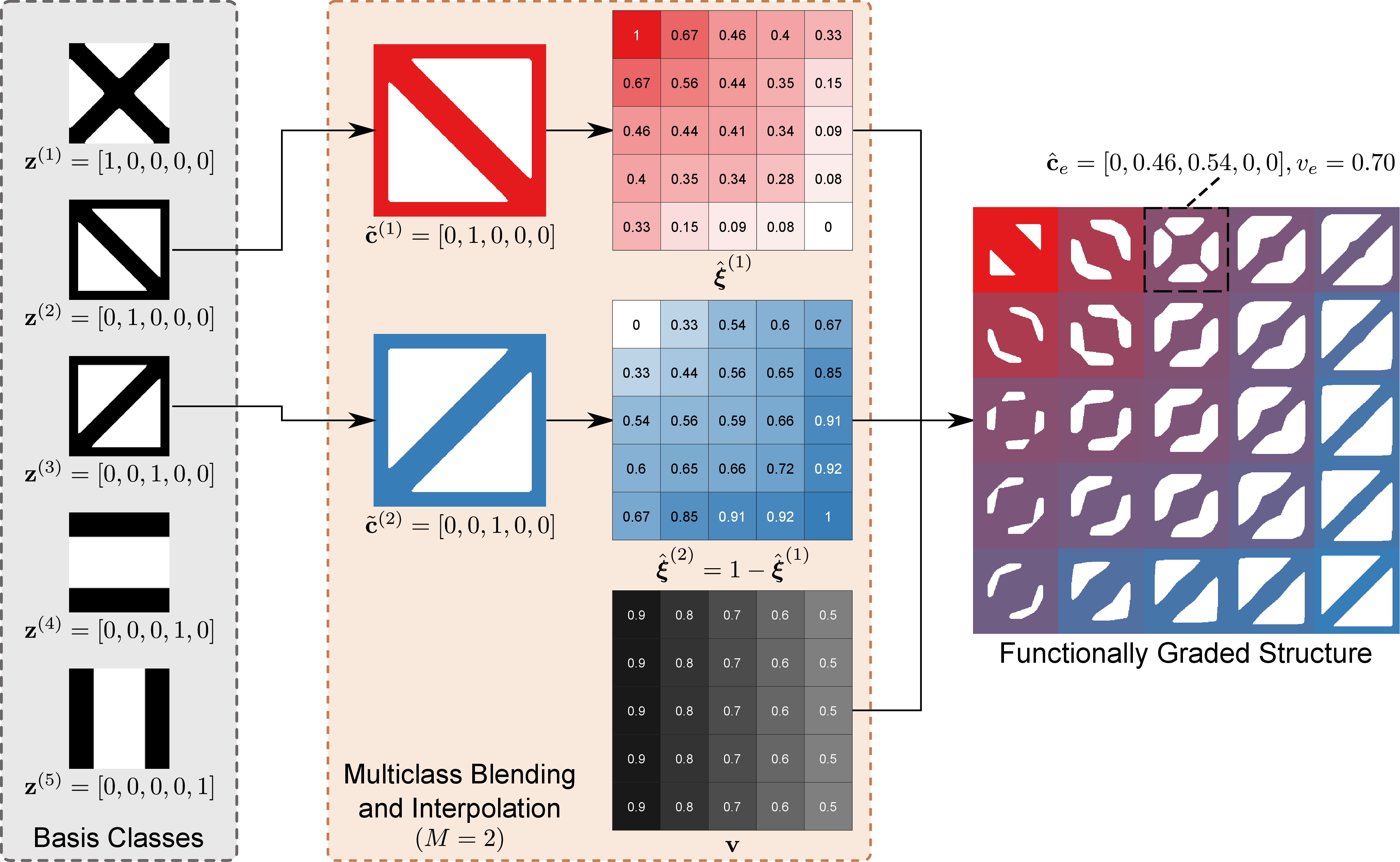}
    \caption{ Demonstration of the integration of multiclass shape blending and global interpolation. The basis classes are defined (left). Examples of $M=2$ new classes, $\tilde{\mathbf{c}}$, their distributions, $\hat{\boldsymbol{\xi}}$, and volume fractions, $\mathbf{v}$, are given (middle). Each microstructure in the FGS (right) is generated by using Eq.~\ref{eq:blend_final}. }
    \label{fig:blend_demo}
\end{figure*}

In the context of multiscale design, the subscript $e$ is added to denote individual microstructures, which each resides in one macroscopic quadrilateral $4$-node finite element. Instead of directly optimizing $D$ weights at each microstructure, we reduce the number of design variables by optimizing $M$ new classes and interpolating them throughout the global structure using the distribution fields $\boldsymbol{\xi}^{(p)}$. An example with $M=2$ is portrayed by the red and blue classes in the middle panel of Fig.~\ref{fig:blend_demo}. For demonstration, we artificially create the values of the macro- and micro-scale design variables in the middle panel, but during design, these values are optimized concurrently.

Similar to multi-material TO (and Eq.~\ref{eq:class_interp}), we require that the sum of the distributions at each element equals one. Thus, to obtain each microstructure, $e$, we can globally interpolate the optimal class weights, $\tilde{\mathbf{c}}^{(m)}$, with:
\begin{equation}\label{eq:global_interp}
    \hat{\mathbf{c}}_e = \tilde{\mathbf{c}}^{(1)} + \sum^{M-1}_{j=1} \bigg[ \big(\tilde{\mathbf{c}}^{(j+1)} - \tilde{\mathbf{c}}^{(j)}\big) \prod^{j}_{k=1} \hat{\xi}^{(k)}_e \bigg],
\end{equation}
where $\hat{\boldsymbol{\xi}}$ are the smoothed distribution fields after applying a radial filter~\citep{sigmund2007filters} to encourage functional grading. 

Therefore, combining Eqs.~\ref{eq:blend_inner} through~\ref{eq:global_interp}, our final multiclass shape blending scheme for a microstructure at element $e$ is:
\begin{equation}\label{eq:blend_final}
\begin{aligned}
    \boldsymbol{\Phi}_e = \frac{1}{\beta_2} \log \Big\{     &
    \exp{\big[ \beta_2 \big( \sum_d^D \hat{c}_{e \cdot d} \boldsymbol{\Phi}_d^{*} + t_e \big) \big]} \\     &
    + \sum_d^D a_{e \cdot d} \exp{\big( \beta_2 \boldsymbol{\Phi}_d^{L} \big)} \Big\},
\end{aligned}
\end{equation}
where $a_{e \cdot d}=H(\hat{c}_{e \cdot d})$ and $t_e$ is found using the bisection algorithm to match a given or optimized volume fraction, $v_e$. This replaces Eq.~\ref{eq:blend_outer} during optimization.

\subsection{Property Prediction with Neural Networks}\label{sec:method_nn}
The continuous and low-dimensional microstructure representation lends itself well to one of the simplest deep learning (DL) techniques: regression with neural networks. More specifically, we can create feedforward neural networks with three or fewer hidden layers that predict the components of a microstructure's effective stiffness tensor, $\mathbf{C}^H_e$, given the scalar values of the blending weights and volume fraction as inputs. That is, $\mathbf{C}^H_e=NN(\hat{\mathbf{c}}_e,\hat{v}_e)$, where $\hat{\mathbf{c}}_e$ and $\hat{v}_e$ are the interpolated class and filtered volume design variables, respectively.

After each hidden layer, we use a $\tanh$ activation function. For training, we use the mean squared error (MSE) loss and the Levenberg-Marquardt optimizer~\citep{Mor1978levenberg}. To include all possible microstructures, such as those where only a few basis classes have non-zero weights, we use an optimal sliced Latin hypercube method~\citep{Ba2015slicedOLHS} to sample the weights $c_d$, first creating combinatorial "slices", then $20$ space-filling samples for each "slice". To cover volume fractions, we also sample 15 microstructures from each resulting SDF (i.e., each set of weights) over $t\in[-1,1]$. We obtain a total of $22,575$ microstructures for $D=5$ basis classes. The effective stiffness tensors of each are calculated using an energy-based homogenization method~\citep{Andreassen2014hom2d}. We set aside $70\%, 15\%, 15\%$ of the data for training, validation and testing. 

Once the model is trained, we can use backpropagation~\citep{Hastie2009learning} to analytically derive the gradients of $\mathbf{C}^H_e$ with respect to the design variables, $\hat{\mathbf{c}}_e$ and $\hat{v}_e$. This, together with the rapid predictions that bypass the cost of homogenization, allows the neural networks to significantly boost the efficiency of design.

\subsection{Concurrent Multiclass Data-Driven Topology Optimization}\label{sec:method_design}
One challenge in creating a concurrent functionally graded design framework that produces realistic results while remaining as general as possible is the different feasible ranges of arbitrary microstructure classes. Consider the five truss basis classes in Sec.~\ref{sec:method_blending}. For a prescribed minimum feature size of $4$ pixels, the first basis has a minimum feasible volume fraction of $0.2$, while the second has a minimum of $0.4$. The question that arises is: when different $v_{min}>0$ are possible for each element $e$, how can we design the distribution of volume fractions while also optimizing a clearly defined macrostructure where some microstructures are allowed to be void (i.e., $v_e=0$)?

Regarding this, most existing FGS research have either elected to ignore the macroscale design altogether~\citep{Li2019tpms}, or adopted a hybridized method that splits the macro- and micro-scale designs into two optimization problems~\citep{Wang2018gradedmorph,Jansen2020hybridfgs,Chu2019fixconn,Zhang2018krigingFGScomp}. None of these works incorporate multiclass designs where the basis topologies can be drastically different, however.

We propose to overcome this hurdle by merging parametric and non-parametric methods in a framework that utilizes evolutionary TO (BESO~\citep{Huang2007beso}) to optimize the discrete global structure, $\mathbf{x}$, and gradient-based TO solved by the method of moving asymptotes (MMA)~\citep{Svanberg1987mma} to concurrently design the coefficients of $M$ new classes $\mathbf{c}$, their distributions $\boldsymbol{\xi}$, and volume fractions $\mathbf{v}$. The approach is similar to the latter group above, but unlike many, we evolve the designs at both scales in the same iteration. This combination allows arbitrary sets of basis microstructures to be used rather than strict constraints or careful handpicking, and distinguishes our framework in terms of generality and efficiency.

Thus, the general optimization problem is:
\begin{equation}\label{eq:to_general}
\begin{aligned}
    & \underset{\mathbf{c},\mathbf{v},\boldsymbol{\xi},\mathbf{x}}{\text{minimize}}
    & & f = f_{perf}(\mathbf{c},\mathbf{v},\boldsymbol{\xi},\mathbf{x}) + kf_{div}(\mathbf{c}),\\
    & \text{subject to}
    & & \mathbf{K}\mathbf{U} = \mathbf{F},\\
    & & & g_j \leq 0, \\
    & & & 0 \leq c_d^{(m)} \leq 1, \\
    & & & 0 \leq v_{min} \leq v_e \leq v_{max}, \\
    & & & 0 \leq \xi_e^{(p)} \leq 1, \\
    & & & x_e \in \{x_{min},1\}, \\
    & & & j \in [1,N_{con}], \quad e \in [1,N_{el}], \\
    & & & m \in [1,M], \quad d \in [1,D-1], \\
    & & & p \in [1,M-1], 
\end{aligned}
\end{equation}
where $f_{perf}$ is an application-dependent measure of design performance, $f_{div}$ is a penalty on low class diversity (Sec.~\ref{sec:method_design_div}), $N_{con}$ is the number of constraints $g_j$ (if any), $N_{el}$ is the number of macroscopic elements or microstructures, and $M$ and $D$ are the numbers of new (to be optimized) and basis (fixed) classes, respectively. A small number, $x_{min}=1$e$-9$, is used to indicate void microstructures to avoid numerical issues. The minimum volume fraction, $v_{min}$, is dependent on the chosen set of basis classes, i.e., $\min(volume(\boldsymbol{\Phi}^L_d))$, whereas the upper bound, $v_{max}$, is $0.95$.

We employ the traditional radial averaging filter~\citep{sigmund2007filters} on our global-level design variables to avoid mesh dependency, resulting in the smoothed fields $\hat{\mathbf{v}}$, $\hat{\boldsymbol{\xi}}^{(p)}$, and $\hat{\mathbf{x}}$. This additionally enforces the interface between optimal classes to be functionally graded, and that the macrostructure has a minimum feature size of $r_{min}$.

To update the designs, we use the default algorithms for BESO~\citep{Huang2007beso} and MMA~\citep{Svanberg1987mma}. The only difference is that, to connect the two scales, the sensitivity number for BESO are dependent on the predicted effective properties of the microstructures. The derivations for this and all other sensitivities in our problem are shown in Appendix~\ref{sec:appen_sens}.

In total, then, our multiclass FGS design framework has $(D-1)M + (M+1)N_{el}$ variables, where $M \leq D \ll N_{el}$.
For the $M=2$ MBB beam example later (Fig.~\ref{fig:ex_mbb1_bcs}), our method has $1,928$ design variables. The multilattice approach by \cite{Sanders2021multilattice} uses $3,200$ variables for the same problem without optimizing the graded volume fractions, while the latent variable multiclass approach of \cite{Wang2020lvgp} has $1,920$ variables and includes functionally graded volumes; both, however require predefined classes with manually defined connections. 
It is important to note that although it is possible for other methods to contain less design variables, they make simplifications that we do not.

\subsubsection{Penalty to Encourage Convergence to Diverse Classes}\label{sec:method_design_div}
Our method can enable high design freedom even with a low-dimensional microscale representation since blending allows the basis classes to mix continuously at both scales. Depending on the chosen optimizer, managing such complexity in two-scale design can be a challenge, one that is also encountered by existing multiscale methods. However, we propose that a cost-effective penalty on the objective function can aid the optimizer (in our case, MMA) without resorting to user-defined restrictions on the design space.

We introduce a penalty on low diversity between the $M$ new classes, encouraging the microscale design variables, $\mathbf{c}^{(m)}$, to converge to values away from each other, so that the FGS is more likely to include different basis classes:
\begin{equation}\label{eq:diversity_pen}
    f_{div} = - \log \big[ \det ( L_{ij}(\mathbf{c}^{(i)},\mathbf{c}^{(j)}) ) \big],
\end{equation}
where $L_{ij}=\exp(-0.5~\lVert \mathbf{c}^{(i)}-\mathbf{c}^{(j)} \rVert_2^2)$ and $i,j\in[1,M]$. 

This is based on determinantal point processes (DPPs), which measure the diversity of a set of items (e.g., the classes $\mathbf{c}^{(m)}$ here) using a similarity matrix $L_{ij}$, whose elements are the similarities between $i$-th and $j$-th pairs of data. The diversity can then be defined as the determinant of $L_{ij}$. A larger determinant value indicates that a set contains less similar items, spans a larger volume, and hence has greater diversity. 
A deeper dive into the benefits of diversity for data-driven multiscale design, can be found in our previous work~\citep{Chan2020metaset}.

Intuitively, minimizing $f_{div}$ is equivalent to maximizing the diversity of the new classes. Since the value of $f_{div}$ approaches zero as classes become more diverse, i.e., the values of $c_d^{(i)}$ and $c_d^{(j)}$ grow farther apart, diversity serves as a natural penalty function. It needs only a weight $k$ so that its value, typically within $[0,1]$ after the first few iterations, can compete with the structural performance, $f_{perf}$. Moreover, it acts similarly to an $L_2$ regularizer that smooths the objective function as $k$ increases, which may avoid sensitivity to initializations and help find an optimum faster in some non-convex or highly nonlinear problems. Indeed, we find in our case studies that adding the penalty help our optimizers to find more optimal solutions (Sec.~\ref{sec:ex_mbb}).

\subsubsection{Volume Relaxation and Adaptive Target Volume}\label{sec:method_design_vf}
If any volume constraints are defined in the design problem, we must have a way to obtain the continuous gradients of volume with respect to the parameters of the shape blending scheme. To achieve this, we can approximate the filtered volume fraction of a microstructure, $\hat{v}_e$, by transforming its SDF into a relaxed grayscale field, similar to that of density-based TO, using the Sigmoid function
\begin{equation}\label{eq:sigmoid}
    S(\boldsymbol{\Phi},\beta_1) = \frac{1}{1+\exp{( -\beta_1 \boldsymbol{\Phi} )}},
\end{equation}
where $\beta_1$ is a fixed parameter to control the strength of relaxation. Thereafter, the approximate volume is
\begin{equation}\label{eq:sig_vf}
    \hat{v}^a_e= \frac{1}{n_{el}} \sum_{u=1}^{n_{el}} S(\Phi_{e\cdot u}, \beta_1),
\end{equation}
where $n_{el}$ is the number of elements in the discretized SDF, and the sensitivity of $\partial V_{Global} / \partial \hat{\mathbf{c}}_e$ is straightforward to calculate (see Appendix~\ref{sec:appen_sens}).

In addition, low volumes are often a goal in multiscale design to take advantage of the porosity of the microstructures. By immediately applying a strict volume fraction constraint, however, it is possible to encounter infeasible and broken structures early in the optimization process~\citep{Christiansen2013explicitTO}. To avoid this, and to ensure that our macro- (BESO) and micro-scale (MMA) designs evolve at approximately the same rate, we use an adaptive scheme to lower the target volumes every $10$ iterations. 
The algorithms of this scheme as well as our complete concurrent design framework can be found in Appendix~\ref{sec:appen_algs}.

\begin{figure*}[h!]
    \centering
    \includegraphics[width=0.8\textwidth]{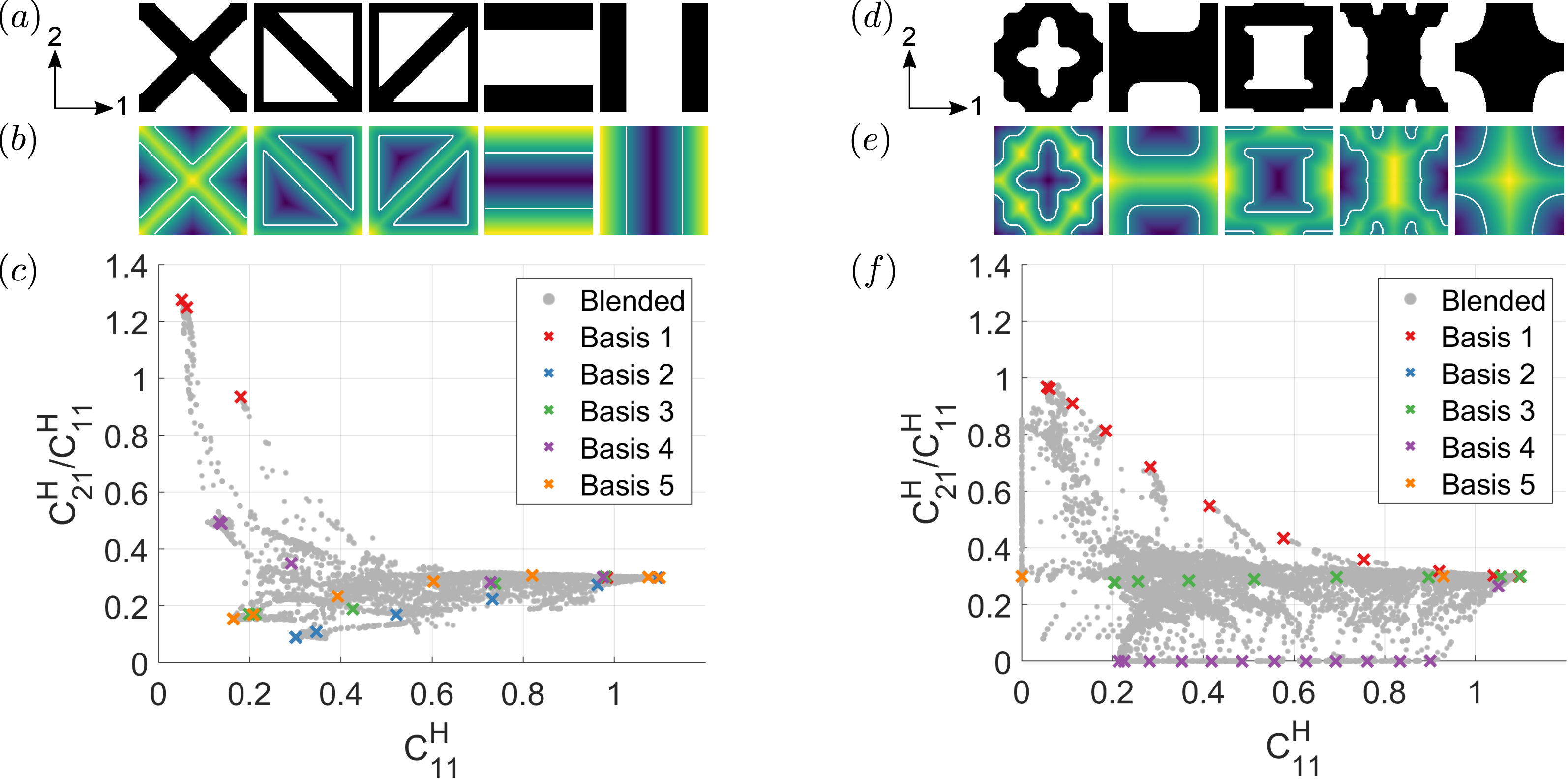}
    \caption{(a) Truss basis classes represented as (b) SDFs ($\boldsymbol{\Phi}^*_d$), and (c) the property space of 22,575 blended microstructures by sampling $\{\mathbf{c},v\}$. (d) Freeform basis classes, (e) their SDFs, and (f) property space.}
    \label{fig:blend_propspace}
\end{figure*}

\begin{table*}[!h]
\begin{center}
\caption{Neural network architectures and accuracies.}\label{table:2d_nn}%
    \begin{tabular}{@{}l ll lll@{}}
    \toprule
                    & $n_{resp}$      & $n_{node}$ & Train $R^2$ (MSE) & Val. $R^2$ (MSE) & Test $R^2$ (MSE) \\ \midrule
    Truss           & 6 & \{16,16,12\}    & 0.9983 (2.34e-4)    & 0.9984 (2.80e-4)     & 0.9983 (2.46e-4)   \\
    Freeform        & 4 & \{12,12,6\}     & 0.9991 (2.91e-4)    & 0.9990 (2.90e-4)     & 0.9991 (3.01e-4)   \\ \bottomrule
    \end{tabular}
\footnotetext{Val.: Validation.}
\end{center}
\end{table*}

\section{Illustrative Examples}\label{sec:examples}
Through several linear elastic problems, we test the ability of our framework to achieve smooth and feasible functional grading of microstructure morhpologies. Namely, we design two compliance and one shape morphing examples. For each case, we study the effects of two sets of basis classes with different morphology types and initial mutual compatibility, as well as the number of new optimal classes, $M$. 

\subsection{Basis Classes and Neural Networks}\label{sec:ex_select}
To illustrate the framework across a range of microstructure morhpologies commonly found in literature, we use two sets of basis classes: one consisting of trusses, and one of topology-optimized freeform shapes. Moreover, to show that human bias can be removed from the design without sacrificing much performance, we compare handpicking the truss basis to automatically selecting the freeform basis using diversity metrics.

\subsubsection{Handpicked Simple Trusses}\label{sec:ex_select_truss}
Truss-type microstructures possess both simple definitions and satisfactory performance~\citep{Panetta2015,Wang2020lvgp,Luo2021fixinterclass,Chan2019multimat}. As such, they are fitting basis classes to validate the proposed framework. For our examples, we choose a set that can, when combined in various ways, cover nearly all of the common truss morphologies in literature. Departing from other methods, we dial up the difficulty by defining the last two bases so that they are broken; to obtain feasible designs, these classes need to be either well-connected to their neighbors or blended with other bases that have self-connectedness. The five classes are shown in Fig.~\ref{fig:blend_propspace}a and~b.

\subsubsection{Shape and Property Diverse Freeform Subsets}\label{sec:ex_select_diverse}
We also assess the efficacy of our blending and interpolation schemes under even more challenging circumstances by defining a set of freeform basis classes with complex shapes derived from TO (Fig.~\ref{fig:blend_propspace}d). They present interesting and highly illustrative case studies as their compatibility with each other is quite low. If used directly in design without our shape blending scheme, the feasibility of the final design would be extremely challenging to guarantee.

We collect these freeform classes from a different perspective, one where a designer has little prior knowledge and wishes to avoid using costly inverse optimization to find the basis microstructures. 
Thus, the five freeform classes are chosen by leveraging the open-source 2D metamaterials dataset~\citep{Wang2020smo} and the automated diverse subset selection method~\citep{Chan2020metaset} from our previous works. The method utilizes the DPPs introduced earlier (Sec.~\ref{sec:method_design_div}) to maximize the shape and property diversity of a subset of microstructures. By automatically covering a wide range of shapes and properties, we hypothesize that diverse basis classes can provide a high return on investment, attaining competent or even superior performance across multiple applications with less effort during the selection of bases.

We filter out any $50 \times 50$ microstructures with minimum feature sizes less than $4$ pixels prior to applying our subset selection method. This eliminates some of the most complex shapes that provide little benefit for functional grading (e.g., microstructures with thin features that would limit the range of feasible volume fractions). To convert the selected binary microstructures into continuous SDF representations, we use the fast marching method~\citep{sk-fmm}. 
The shape and property diverse freeform basis classes shown in Fig.~\ref{fig:blend_propspace}d and~e.

\subsubsection{Property Prediction Models}\label{sec:ex_models}
As discussed in Sec.~\ref{sec:method_nn}, we obtain training datasets of $22,575$ microstructures using sliced Latin hypercube sampling for each set of basis classes. 
The respective property spaces are depicted in Fig.~\ref{fig:blend_propspace}c and~f, where samples generated from blending are in gray and those from the original basis classes are denoted as different color crosses. 

We observe that the blended microstructures are able to interpolate between~--~and in some cases, extend slightly beyond~--~basis classes in order to cover the property space. Although some sparse areas still exist due to the lower feasible bounds that we impose, this shows that blending is a powerful technique to create a large design space even with small sets of basis classes. 
It is also clear from these figures that the first basis (red) from both sets possess the highest stiffness in diagonal directions and ratios of $C^H_{21}$ to $C^H_{11}$ (equivalent to the effective Poisson's ratio), as opposed to truss bases 4 (purple) and 5 (orange), and freeform basis 2 (blue), which are stiffest in uniaxial directions. 

While the freeform classes are orthotropic, some of the truss classes are not. The number of responses, $n_{resp}$, in the neural networks of each set are adjusted accordingly, i.e., 6 elastic tensor components for trusses and 4 for freeform. Table~\ref{table:2d_nn} lists the details of the ML model architectures, where $n_{node}$ indicates the neurons of each hidden layer, and the $R^2$ and MSE metrics of the trained models. 

Since they only need to be built once, the same models are used for all examples throughout the paper.
The one-time expense of creating our data and models is reasonable for our 2D problems. However, we note that our results show overlapping properties in the dataset (Fig.~\ref{fig:blend_propspace}c and f), high $R^2$ values above $0.99$ and low MSE (Table~\ref{table:2d_nn}), suggesting that it may not have been necessary to use as many samples as we did. There is great potential to develop adaptive sampling algorithms that better balance accuracy and efficiency, particularly for 3D and complex applications. We leave this for future works.

\subsection{Compliance Minimization}\label{sec:ex_mbb}
\begin{figure}[th!]
    \centering
    \begin{subfigure}[c]{0.85\columnwidth}
        \centering
        \includegraphics[width=1\columnwidth]{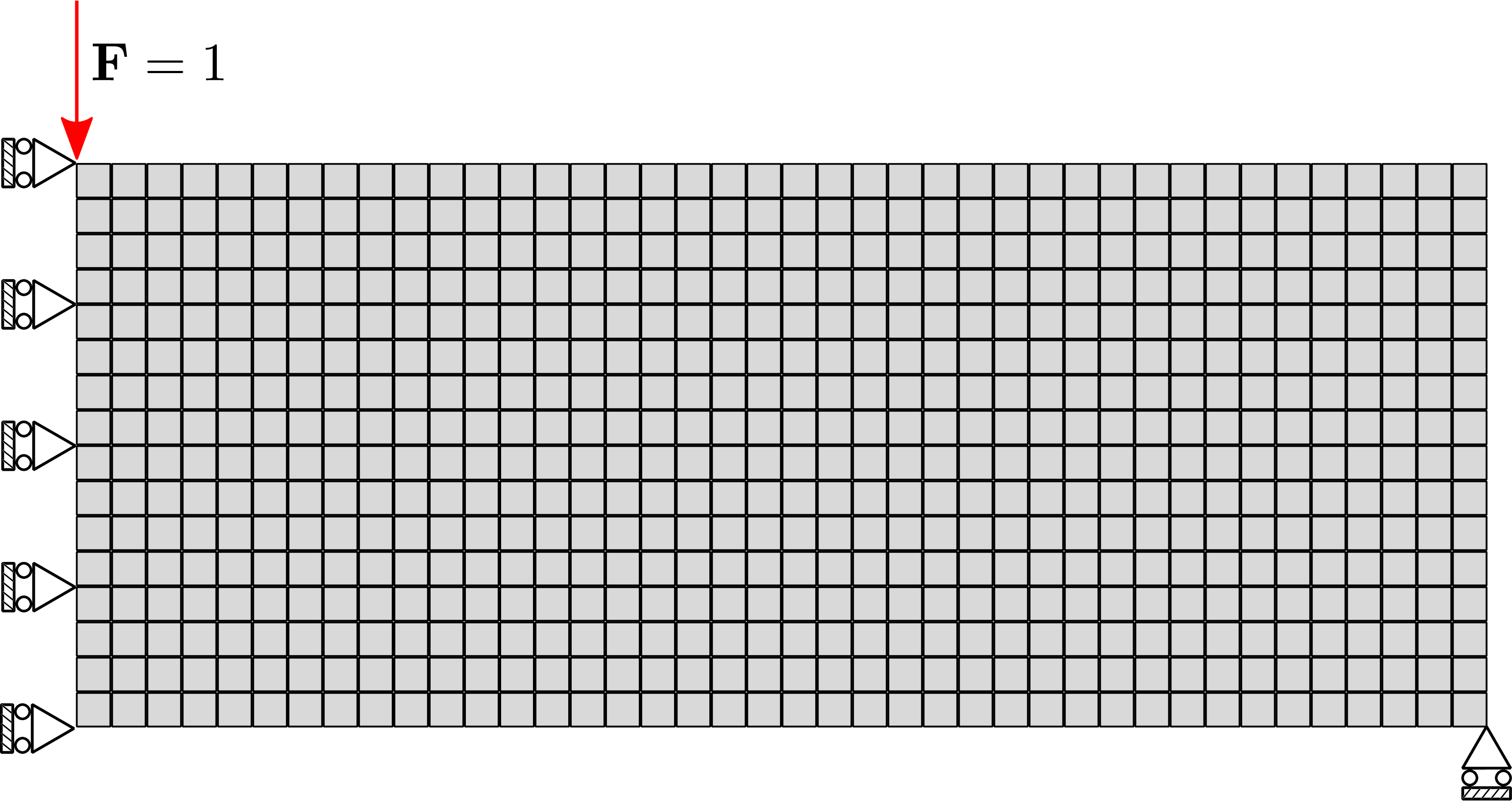}
        \caption{MBB beam ($40\times16$)}
        \label{fig:ex_mbb1_bcs}
    \end{subfigure} %
    \begin{subfigure}[c]{0.85\columnwidth}
        \centering
        \includegraphics[width=1\columnwidth]{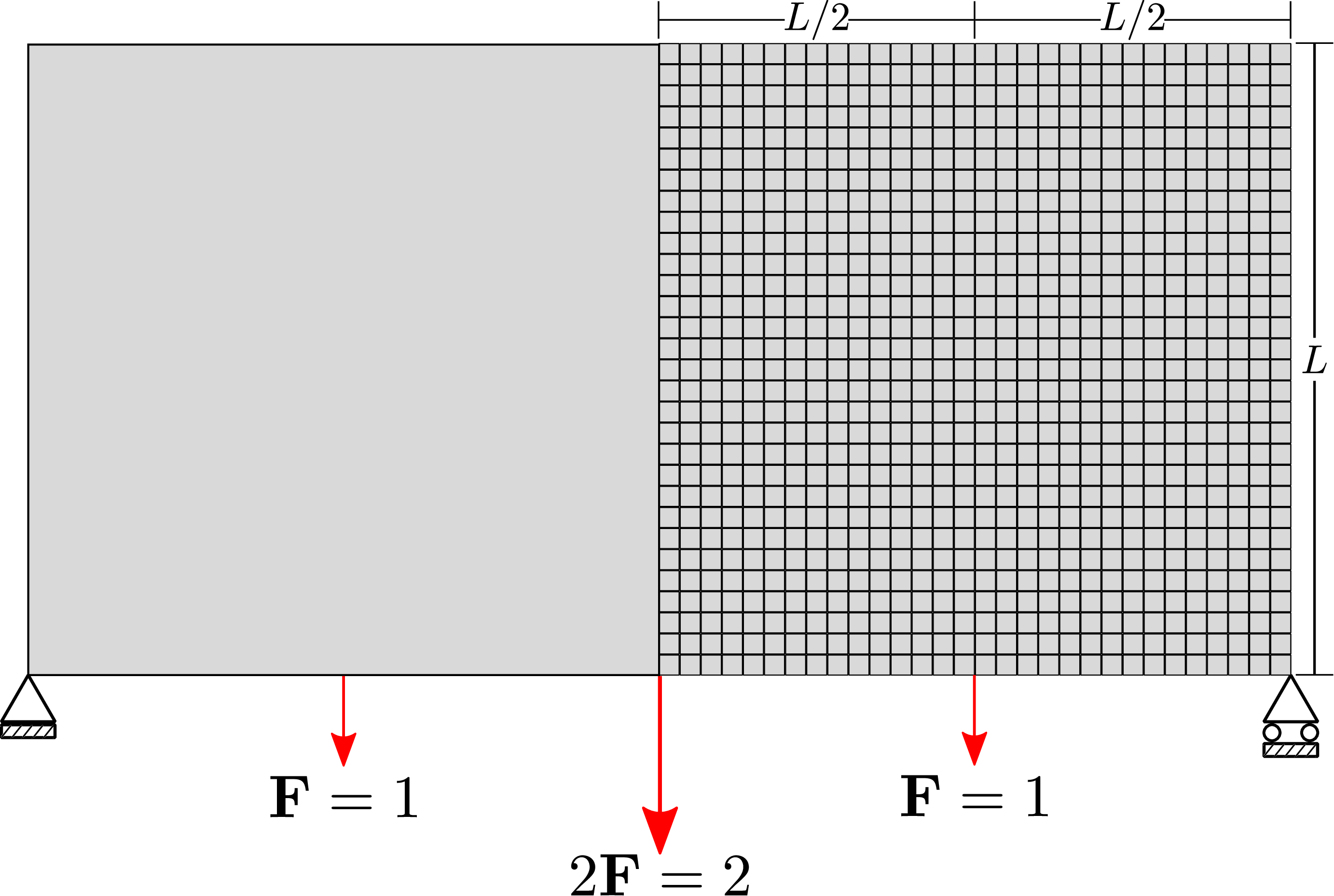}
        \caption{Bridge ($30\times30$)}
        \label{fig:ex_mbb2_bcs}
    \end{subfigure}
    \caption{Problem settings of the compliance minimization examples.}
    \label{fig:ex_mbb_bcs}
\end{figure}

We begin with compliance minimization examples, the first of which is the classic MBB beam. The boundary conditions are depicted in~\ref{fig:ex_mbb1_bcs}, and we follow the same set-up as in~\cite{Xia2014FE2} and~\cite{Wang2020lvgp} in order to compare our results with those of existing methods. That is, the MBB beam is discretized into $40\times 16$ microstructures and an ambitious global volume fraction limit is set as $V_{Global}^*=0.36$. 
For the second problem, we pursue a $60\times 30$ bridge structure loaded in three places, as shown in Fig.~\ref{fig:ex_mbb2_bcs}. Due to symmetry, we can cut its size by half into $30 \times 30$. The target global volume there is $V_{Global}^*=0.50$.

The two compliance problems can be formulated as:
\begin{equation}\label{eq:to_compliance}
\begin{aligned}
    & \underset{\mathbf{c},\mathbf{v},\boldsymbol{\xi},\mathbf{x}}{\text{minimize}}
    & & f = f_c + kf_{div}(\mathbf{c}),\\
    & \text{subject to}
    & & \mathbf{K}\mathbf{U} = \mathbf{F},\\
    & & & g_1 = V_{Global} / V^*_{Global} - 1 \leq 0,\\
    & & & g_2 = V_{BESO} / V^*_{BESO} - 1 \leq 0,
\end{aligned}
\end{equation}
where $f_c=\sum_{e=1}^{N_{el}} \mathbf{u}_e^T \mathbf{k}^H_e(\hat{\mathbf{c}}_e,\hat{v}_e) \mathbf{u}_e$ is the compliance with element displacements $\mathbf{u}_e$ and effective stiffness matrices $\mathbf{k}^H_e$, which are predicted via the DL models. The global and macroscale volumes are $V_{Global} = \sum_{e=1}^{N_{el}} x_e \hat{v}_e / N_{el}$ and $V_{BESO} = \sum_{e=1}^{N_{el}} x_e / N_{el}$, respectively, and the bounds on the design variables are the same as described in Eq.~\ref{eq:to_general}. The sensitivities of this optimization problem are detailed in Appendix~\ref{sec:appen_sens}.

For both examples, we initialize the volume constraints as $V^*_{Global,0}=0.95, V^*_{BESO,0}=\sqrt{V^*_{Global}}$, the volumes $\mathbf{v}$ as $0.95$, the class weights so that $\tilde{\mathbf{c}}^{(m)}$ are the same, and all distribution fields, $\boldsymbol{\xi}^{(p)}$, so that the classes are distributed equally. The penalty parameter (Sec.~\ref{sec:method_design_div}) is set so that $kf_{div}=10$ during later iterations. We find that keeping the penalty around this value improves both convergence and design performance. 

The fixed parameters in the shape blending scheme are $\beta_1=64$ and $\beta_2=32$. As we suggested in Sec.~\ref{sec:method_blend_blend}, the threshold $\eta_2$ is adapts to the design and is equal to the $75th$-percentile of the class weights. The radii of all filters on the design variables are the same, $r_{min}=3.0$, matching \cite{Xia2014FE2}. 
For BESO, the evolutionary rate is $ER=0.05$ in both problems. Otherwise, all other parameters are kept at the default values~\citep{Huang2007beso,Svanberg1987mma}. 
Our convergence criteria are when the change in design or the mean change in the objective over $10$ iterations are less than $0.01$, or when the number of iterations reaches $200$. We also use early-stopping if the target $V_{Global}^*$ has been met but the objective has not improved in $20$ iterations.

In the following sections, we present our results in figures with the same layout: The left sides illustrate how the optimized classes are created via blending by breaking them down into the individual basis classes. For ease of interpretation, we show $\tilde{c}_d^{(m)}$ from Eq.~\ref{eq:class_interp}, which correspond directly to the weights used during blending (Eq.~\ref{eq:blend_final}), instead of the design variables $c_d^{(m)}$. 
The right-most sides show the optimal multiclass FGS and their final compliance.

For fair comparison between our results and homogenization-based methods in literature, and between the two types of basis morphologies, all compliance values stated in the main paper are calculated using numerical homogenization. 
For further validation, Tables~\ref{tab:comp_compare_mbb} and~\ref{tab:comp_compare_bridge} in Appendix~\ref{sec:appen_results} also reports the compliance obtained from the neural networks and fine mesh analysis.

\subsubsection{2-Class Results with Different Basis Classes}\label{sec:ex_mbb_M2}

\begin{figure*}[th]
    \centering
    \includegraphics[width=0.9\textwidth]{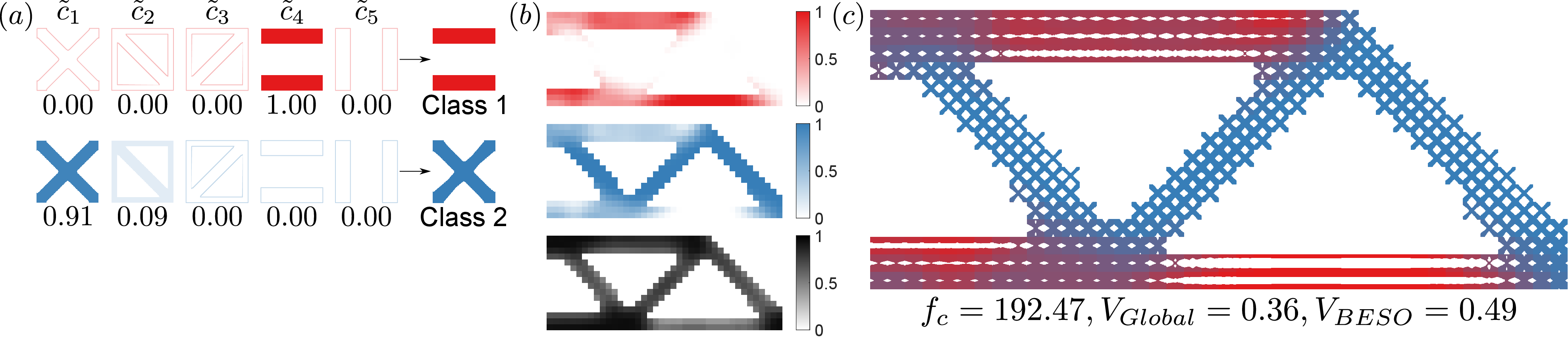}
    \caption{Truss MBB, 2-class result. (a) Optimal new classes each drawn in a different color. Left of arrows: optimal weights listed under each basis. Lighter colors indicate low weights while outlined shapes represent weights that are zero. Right of arrows: representative topologies of new classes. (b) $\hat{\boldsymbol{\xi}}^{(1)},\hat{\boldsymbol{\xi}}^{(2)},\hat{\mathbf{v}}$ from top to bottom, and (c) multiclass FGS.}
    \label{fig:ex_truss_mbb1_2class}
    \includegraphics[width=0.9\textwidth]{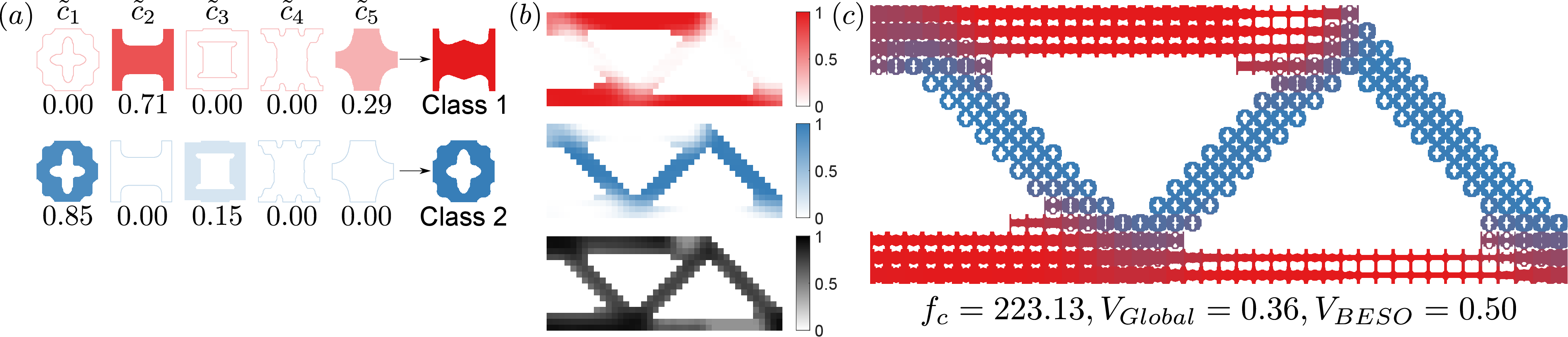}
    \caption{Freeform MBB, 2-class result: optimal (a) weights and representatives of new classes, (b) $\hat{\boldsymbol{\xi}}^{(1)},\hat{\boldsymbol{\xi}}^{(2)},\hat{\mathbf{v}}$ from top to bottom, (c) multiclass FGS.}
    \label{fig:ex_dpp_mbb1_2class}
\end{figure*}

\begin{figure*}[th]
    \centering
    \includegraphics[width=0.8\textwidth]{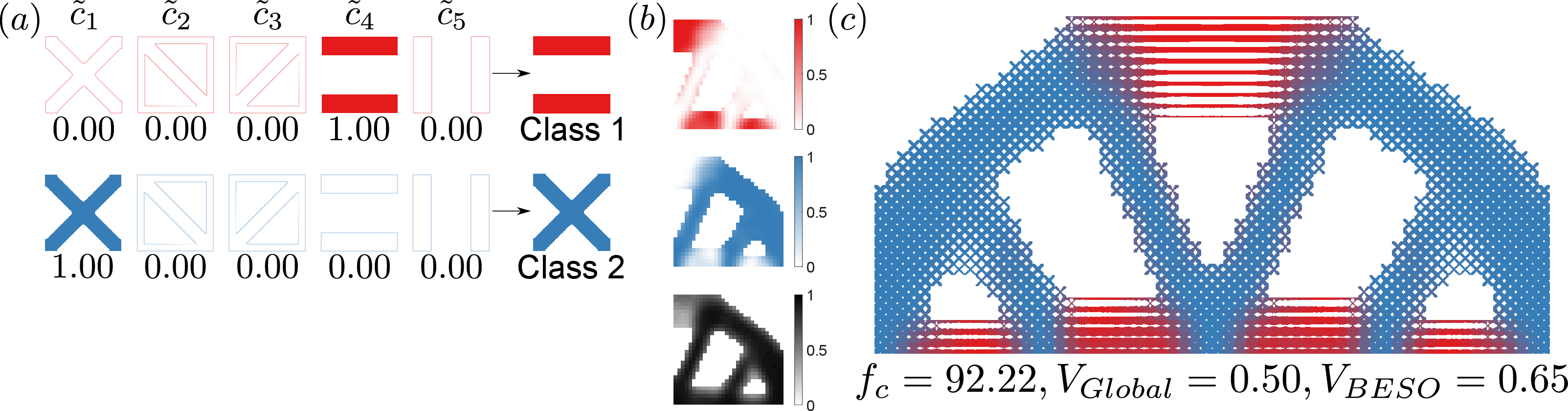}
    \caption{Truss bridge, 2-class result: optimal (a) weights and representatives of new classes, (b) $\hat{\boldsymbol{\xi}}^{(1)},\hat{\boldsymbol{\xi}}^{(2)},\hat{\mathbf{v}}$ from top to bottom, (c) multiclass FGS reflected over the symmetry line.}
    \label{fig:ex_truss_mbb2_2class}
    \includegraphics[width=0.8\textwidth]{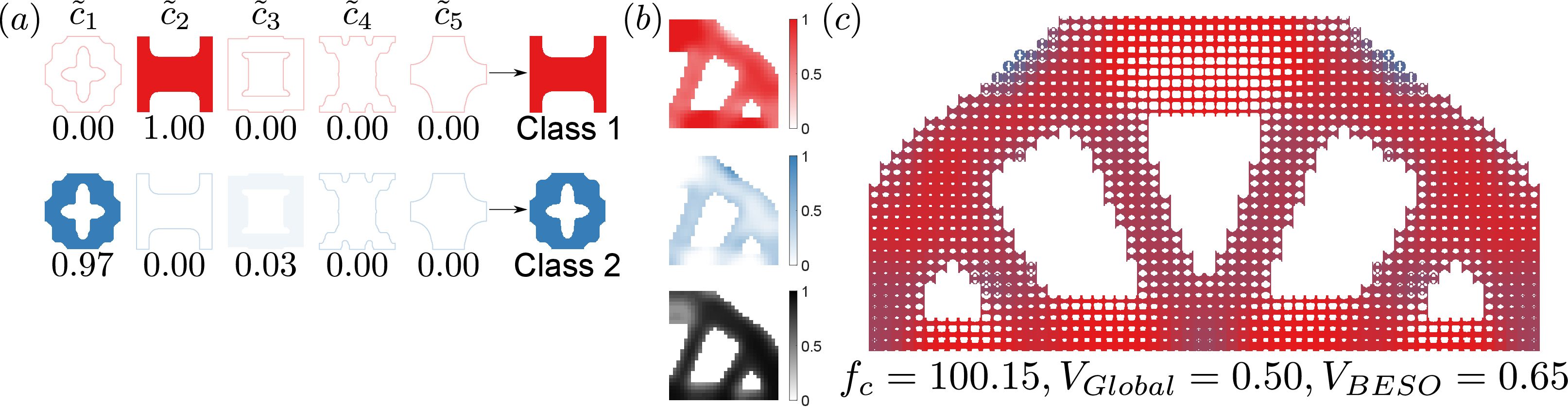}
    \caption{Freeform bridge, 2-class result: optimal (a) weights and representatives of new classes, (b) $\hat{\boldsymbol{\xi}}^{(1)},\hat{\boldsymbol{\xi}}^{(2)},\hat{\mathbf{v}}$ from top to bottom, (c) multiclass FGS reflected over the symmetry line.}
    \label{fig:ex_dpp_mbb2_2class}
\end{figure*}

We first consider the results using $M=2$ new classes and both sets of basis classes. The optimal designs are shown in Figs.~\ref{fig:ex_truss_mbb1_2class} and~\ref{fig:ex_dpp_mbb1_2class} for the MBB beam with truss and freeform bases, respectively, and in Figs.~\ref{fig:ex_truss_mbb2_2class} and~\ref{fig:ex_dpp_mbb2_2class} for the bridge. The interpretation of the figures is described in the previous section.

From the 2-class compliance results, we can see several benefits of integrating multiclass shape blending into FGS design:
\begin{enumerate}
    \item The combination of the blending scheme and the radial filters on the distribution fields creates smooth transitions between classes. Topological functional grading is guaranteed and does not depend on the mutual compatibility of the basis classes. Although connections may not be ideal for our freeform bases, which have low initial connectivity and more complex features, neighboring microstructures change continuously and are at least connected through the imposed lower feasible bounds (Fig.~\ref{fig:ex_dpp_mbb1_2class}). 
    \item Because of the two-step blending scheme, the microstructures at the interfaces of optimized classes are a union of the classes being mixed there, and the minimum feature sizes of all microstructures match our prescribed lower limit of $4$ pixels.
    \item The macroscale distributions, $\hat{\boldsymbol{\xi}}^{(p)}$, can be either dominated solely by one class or contain mixtures of multiple classes. For example, the diagonal struts in the truss MBB (Fig.~\ref{fig:ex_truss_mbb1_2class}a) consist predominantly of the second new class (blue), while the horizontal bars contain both (red and blue), presumably to stiffen the design at those locations. On the contrary, the two new classes in the freeform bridge intermingle throughout nearly the entire structure (Fig.~\ref{fig:ex_dpp_mbb2_2class}). 
    \item Optimizing $c_d^{(m)}$ can automatically determine if an existing basis class is sufficient to achieve low compliance, or if a novel class needs to be created by fusing several bases. For example, in the freeform MBB result (Fig.~\ref{fig:ex_dpp_mbb1_2class}), the mixture of the second and fifth bases stiffens the microstructures, improving the global compliance of the FGS.
    \item Due to BESO, the global macrostructures are clearly defined and change based on the basis classes and optimal microstructures, showing that the hybrid framework works well.
\end{enumerate}

Further observations can be made regarding the framework's ability to adapt to spatially-varying stress distributions. The first and fourth truss-type bases, and the first and second for freeform, are the most popular classes by far, agreeing with our observation in Sec.~\ref{sec:ex_models} that these are among the strongest classes in diagonal and uniaxial directions. In both beam and bridge examples, these classes are designed such that the load-bearing features of the blended microstructures intuitively match the load paths.

In particular, our truss-type MBB beam result is akin to those in existing multiscale works with the same design domain. \cite{Xia2014FE2} performed an exhaustive two-scale TO that optimized every microstructure, resulting in horizontal (uniaxial) and diagonal (anisotropic) features that are oriented with stress directions, and a compliance of $f_c=190$. Meanwhile, \cite{Wang2020lvgp} proposed a multiclass design with rectangular trusses on the horizontal macro-bars, X's on the diagonal macro-struts, and a compliance of $f_c=214.02$. Our framework can be thought of as bridging these two methods. This is indeed reflected in our 2-class truss result, which achieves a compliance value between the two existing works, $f_c=192.47$, and has similar microstructures.

In terms of performance, truss basis classes outshine the freeform basis in both problems. We theorize that our freeform bases do not perform as well for two reasons. (1) They were automatically chosen to maximize diversity, i.e., coverage, in shapes and properties, which undoubtedly can skip microstructures with properties that are more optimal for these specific problems. (2) They contain complex, thin features that force their lower feasible bounds to have high volume fractions ($v_{min}=0.28$), which clashes with the low target volume of $0.36$ in the MBB problem. 

Despite these disadvantages, however, the compliance attained by using freeform basis classes is decent across both examples and near those of existing works, confirming the versatility of diverse bases and our design framework. 
A deeper look into the MBB example also reveals that the multiclass freeform FGS (Fig.~\ref{fig:ex_dpp_mbb1_2class}) still surpasses single-class designs that vary only in volume fraction (Table~\ref{tab:ex_mbb1_single_class}).

\begin{table}[b]
\centering
\caption{Single-class MBB beam results using one freeform basis class each.}
\label{tab:ex_mbb1_single_class}
\begin{tabular}{@{}llllll@{}}
\toprule
Basis & $1$ & $2$ & $3$ & $4$ & $5$ \\ \midrule
$f_c$ & $476.26$ & $247.57$ & $384.89$ & $399.29$ & $290.75$ \\ \bottomrule
\end{tabular}
\end{table}

\begin{figure*}[th]
    \centering
    \includegraphics[width=0.9\textwidth]{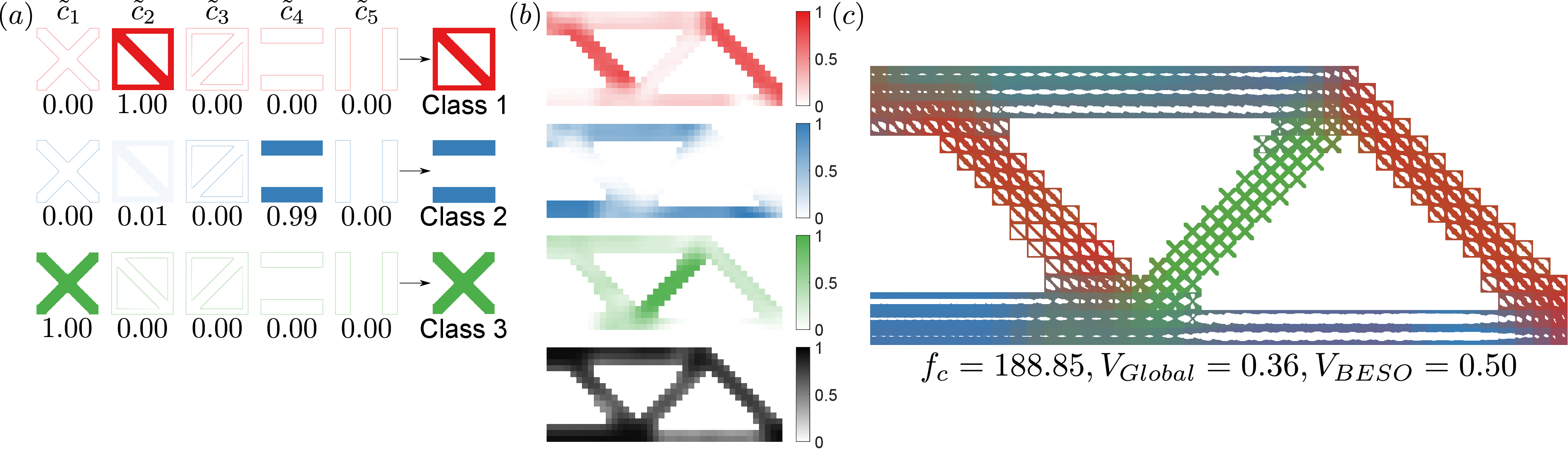}
    \caption{Truss MBB, 3-class result: optimal (a) weights and representatives of new classes, (b) $\hat{\boldsymbol{\xi}}^{(1)},\hat{\boldsymbol{\xi}}^{(2)},\hat{\boldsymbol{\xi}}^{(3)},\hat{\mathbf{v}}$ from top to bottom, (c) multiclass FGS.}
    \label{fig:ex_truss_mbb1_3class}
    \includegraphics[width=0.9\textwidth]{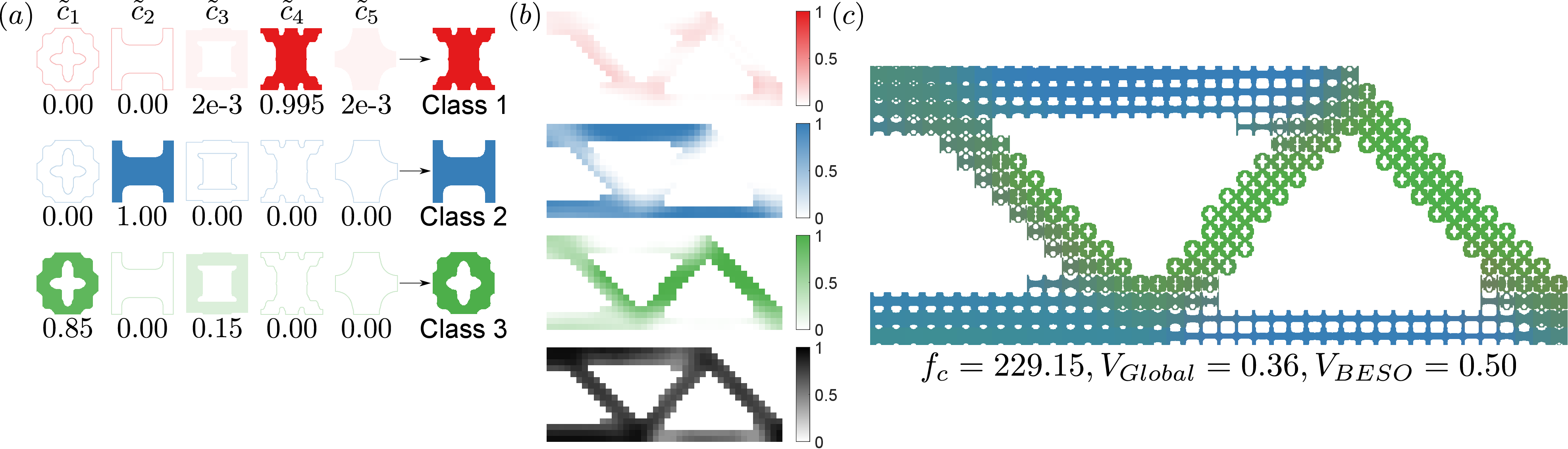}
    \caption{Freeform MBB, 3-class result: optimal (a) weights and representatives of new classes, (b) $\hat{\boldsymbol{\xi}}^{(1)},\hat{\boldsymbol{\xi}}^{(2)},\hat{\boldsymbol{\xi}}^{(3)},\hat{\mathbf{v}}$ from top to bottom, (c) multiclass FGS.}
    \label{fig:ex_dpp_mbb1_3class}
\end{figure*}

\subsubsection{Effect of the Number of New Classes}\label{sec:ex_mbb_M3}
\begin{figure*}[th]
    \centering
    \includegraphics[width=0.74\textwidth]{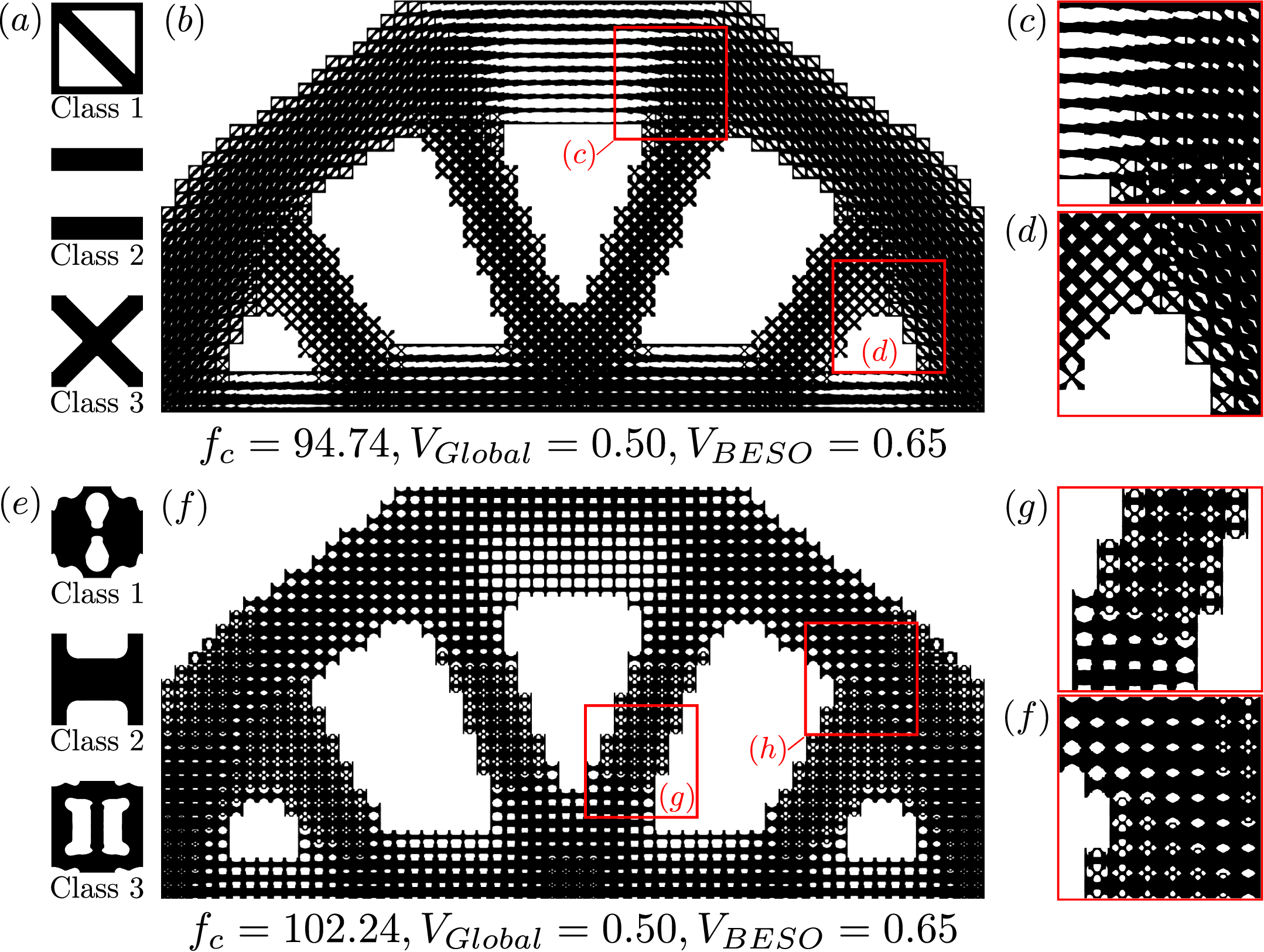}
    \caption{Results of 3-class bridge in black-and-white. For truss bases: (a) optimal classes, (b) FGS, and (c,d) zoom-in views. For freeform bases: (e) optimal classes, (f) FGS, and (g,h) zoom-in views.}
    \label{fig:ex_all_mbb2_3class}
\end{figure*}

With the effectiveness of the proposed framework established, we now study whether increasing the number of new classes to $M=3$ can impact the designs and their performances. The problem definitions remain the same as before. For the MBB beam, the 3-class results are given in Figs.~\ref{fig:ex_truss_mbb1_3class} and~\ref{fig:ex_dpp_mbb1_3class} with the truss and freeform bases, respectively. The bridge results for both sets of bases are combined in Fig.~\ref{fig:ex_all_mbb2_3class}, where they are shown in black-and-white and with zoomed-in views of the functionally graded topologies and volume fractions.

The most notable result is the 3-class truss MBB, which achieves a compliance even lower than the fully optimized design of \citep{Xia2014FE2} at $f_c=188.85$. More apparent than in the 2-class result above, it exhibits directional load-bearing features (Fig.~\ref{fig:ex_truss_mbb1_3class}) such as the left-to-right diagonal microstructures (red) and the near-uniaxial microstructures (blue). In addition, the X-shaped class (green) appears mainly in the middle macro-strut. Overall, our result distinguishes itself from the existing with the mix of isotropic and anisotropic microstructures that are all well-connected. A similar blend of directional and uniaxial classes can be found in our 3-class truss bridge (Fig.~\ref{fig:ex_all_mbb2_3class}a-d). 

With the exception of the truss-type MBB, however, all 3-class results could not overtake the compliance of their 2-class counterparts. We suspect this is because, for simple compliance problems, additional classes are not necessary to achieve optimal performance. Our hypothesis is supported by the new classes of the $M=2$ examples: for the most part, they are each monopolized by just one basis class. This suggests that, in most cases, only two basis classes are needed throughout the entire FGS; if more are required, they can be incorporated into a single optimal class by adjusting the values of $\mathbf{c}^{(m)}$ without increasing $M$, like in the 2-class freeform MBB (Fig.~\ref{fig:ex_dpp_mbb1_2class}). 
Another reason could be that we force the third class to be different from the others through the penalty on low diversity, which can lead to the addition of a sub-optimal class. This scenario may have occurred in the 3-class freeform MBB, where the first new class (red) is hardly present in the FGS. Nevertheless, the penalty does not significantly worsen the compliance and can, in some cases improve it.
An intriguing possibility that bears further investigation is whether a larger variety of microstructures are needed in problems with finer discretization or more complex objectives.

Finally, we compare the computational efficiency of our proposed framework against others, with the caveat that each method was run on different computers. The $40 \times 16$ MBB design is reported to require 200 hours in~\cite{Xia2014FE2}, and 5 minutes in the data-driven method of~\cite{Wang2020lvgp}. For our proposed method, the same design using $M=3$ new classes takes under 12 minutes. However, we note that the majority of this time is consumed by the bisection algorithm (Sec.~\ref{sec:method_blend_blend}), which ensures that the microstructures have the optimized volume fractions when converted from SDFs (Eq.~\ref{eq:sigmoid}). Improving this aspect of our blending scheme is next in our future goals.

\begin{table*}[]
\centering
\caption{MBB beam results without penalty on low class diversity ($k=0$). The dominant weights of each new class, $\Tilde{\mathbf{c}}_m$, are in \textbf{bold}.}
\label{tab:ex_mbb1_no_diversity}
\begin{tabular}{@{}clllll@{}}
\toprule
Basis &
  \multicolumn{1}{c}{$M$} &
  \multicolumn{1}{c}{$\Tilde{\mathbf{c}}_m$} &
  \multicolumn{1}{c}{$f_c$} &
  \multicolumn{1}{c}{$V_{Global}$} & 
  \multicolumn{1}{c}{$V_{BESO}$} \\ \midrule
\multirow{2}{*}{Truss} &
  2 &
  \begin{tabular}[c]{@{}l@{}}$\Tilde{\mathbf{c}}_1=[0.00,\mathbf{0.39},0.00,0.27,0.33]$\\ $\Tilde{\mathbf{c}}_2=[0.00,0.05,0.11,\mathbf{0.84},0.00]$\end{tabular} &
  $205.77$ &
  $0.36$ & $0.50$ \\ \cmidrule(l){2-6} 
 &
  3 &
  \begin{tabular}[c]{@{}l@{}}$\Tilde{\mathbf{c}}_1=[0.00,\mathbf{0.36},0.22,0.12,0.30]$\\ $\Tilde{\mathbf{c}}_2=[0.12,0.12,0.09,\mathbf{0.66},0.01]$\\ $\Tilde{\mathbf{c}}_3=[0.23,0.08,0.16,\mathbf{0.53},0.01]$\end{tabular} &
  $212.83$ &
  $0.36$ & $0.54$ \\ \midrule
\multirow{2}{*}{Freeform} &
  2 &
  \begin{tabular}[c]{@{}l@{}}$\Tilde{\mathbf{c}}_1=[0.00,0.09,0.00,0.00,\mathbf{0.91}]$\\ $\Tilde{\mathbf{c}}_2=[0.04,0.14,0.39,0.01,\mathbf{0.43}]$\end{tabular} &
  $252.02$ &
  $0.36$ & $0.51$ \\ \cmidrule(l){2-6} 
 &
  3 &
  \begin{tabular}[c]{@{}l@{}}$\Tilde{\mathbf{c}}_1=[0.01,0.00,\mathbf{0.53},0.00,0.46]$\\ $\Tilde{\mathbf{c}}_2=[0.00,0.45,0.00,0.00,\mathbf{0.55}]$\\ $\Tilde{\mathbf{c}}_3=[0.00,0.00,0.00,0.00,\mathbf{1.00}]$\end{tabular} &
  $248.84$ &
  $0.36$ & $0.50$ \\ \bottomrule
\end{tabular}
\end{table*}

\subsubsection{Effect of the Low-Diversity Penalty}\label{sec:ex_mbb_div}
The function that penalizes new classes with low diversity (Eq.~\ref{eq:diversity_pen}, Sec.~\ref{eq:diversity_pen}) can affect performance, although whether that effect is positive or negative depends on the problem or basis classes. In this section, we show concrete examples why the penalty is still recommended by running the same compliance problems without the penalty, i.e., by setting $k=0$. The results are listed in Tables~\ref{tab:ex_mbb1_no_diversity} and~\ref{tab:ex_mbb2_no_diversity}. Immediately, we can see that although the truss basis classes can still achieve satisfactory compliance values lower than the existing baselines~\citep{Xia2014FE2,Wang2020lvgp}, none of these can beat our results above. 

In the tables, we write the highest weight values of each new class, $\Tilde{\mathbf{c}}_m$, in bold. From this, we observe that each result is often overshadowed by one basis class (see the bold values in the same column). For the 3-class designs in particular, the second and third new classes are always dominated by the same basis, confirming our earlier suspicion that $M=2$ is enough to produce optimal results. We also note that there are numerous low values of $\Tilde{\mathbf{c}}_m$, signifying that multiple basis classes are being blended into the FGS without improving the design performance. 
These results additionally imply that greater diversity amongst the microstructure classes improves performance. A more meticulous study on the impact of diversity on the generality and performance of design methods is an intriguing path for future works.

\begin{table*}[]
\centering
\caption{Bridge results without penalty on low diversity ($k=0$). The dominant weights of each new class, $\Tilde{\mathbf{c}}_m$, are in \textbf{bold}.}
\label{tab:ex_mbb2_no_diversity}
\begin{tabular}{@{}clllll@{}}
\toprule
Basis &
  \multicolumn{1}{c}{$M$} &
  \multicolumn{1}{c}{$\Tilde{\mathbf{c}}_m$} &
  \multicolumn{1}{c}{$f_c$} &
  $V_{Global}$ & $V_{BESO}$ \\ \midrule
\multirow{2}{*}{Truss} &
  2 &
  \begin{tabular}[c]{@{}l@{}}$\Tilde{\mathbf{c}}_1=[0.00,0.22,0.00,\mathbf{0.73},0.05]$\\ $\Tilde{\mathbf{c}}_2=[0.00,0.41,0.00,0.09,\mathbf{0.51}]$\end{tabular} &
  $96.59$ &
  $0.50$ & $0.65$ \\ \cmidrule(l){2-6} 
 &
  3 &
  \begin{tabular}[c]{@{}l@{}}$\Tilde{\mathbf{c}}_1=[0.00,0.18,0.00,\mathbf{0.82},0.00]$\\ $\Tilde{\mathbf{c}}_2=[0.00,0.28,0.00,0.21,\mathbf{0.51}]$\\ $\Tilde{\mathbf{c}}_3=[0.00,0.45,0.00,0.00,\mathbf{0.54}]$\end{tabular} &
  $96.34$ &
  $0.50$ & $0.65$ \\ \midrule
\multirow{2}{*}{Freeform} &
  2 &
  \begin{tabular}[c]{@{}l@{}}$\Tilde{\mathbf{c}}_1=[0.00,0.00,0.13,\mathbf{0.87},0.00]$\\ $\Tilde{\mathbf{c}}_2=[0.03,0.00,0.06,\mathbf{0.91},0.00]$\end{tabular} &
  $105.90$ &
  $0.50$ & $0.65$ \\ \cmidrule(l){2-6} 
 &
  3 &
  \begin{tabular}[c]{@{}l@{}}$\Tilde{\mathbf{c}}_1=[0.00,0.00,0.10,\mathbf{0.90},0.00]$\\ $\Tilde{\mathbf{c}}_2=[0.00,0.00,0.09,\mathbf{0.91},0.00]$\\ $\Tilde{\mathbf{c}}_3=[0.02,0.00,0.02,\mathbf{0.96},0.00]$\end{tabular} &
  $105.46$ &
  $0.50$ & $0.65$ \\ \bottomrule
\end{tabular}
\end{table*}

Furthermore, it takes significantly longer for the class design variables to converge without penalization. In the 2-class truss bridge example without the penalty, they often fluctuate and need more than $100$ iterations to start converging, whereas they are already converged in under $30$ iterations with penalization. 
These studies validate the benefits that our proposed low-diversity penalty function supply to MMA, helping it to stabilize, escape local minima, and find more optimal solutions. 
In Appendix~\ref{sec:appen_results}, we provide the convergence plots of the $M=3$ MBB beam examples, along with additional discussion.

\subsection{Shape Matching}\label{sec:ex_match_shape}
\begin{figure}[t!]
    \centering
    \begin{subfigure}{0.9\columnwidth}
        \centering
        \includegraphics[width=1\columnwidth]{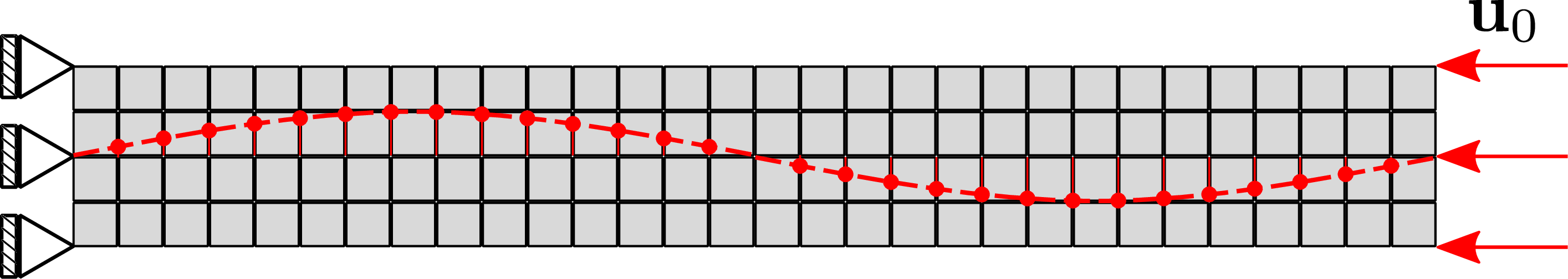}
        \caption{Target sine wave ($30 \times 4$)}
        \label{fig:ex_Uysine_bcs}
    \end{subfigure} %
    \begin{subfigure}{0.9\columnwidth}
        \centering
        \includegraphics[width=1\columnwidth]{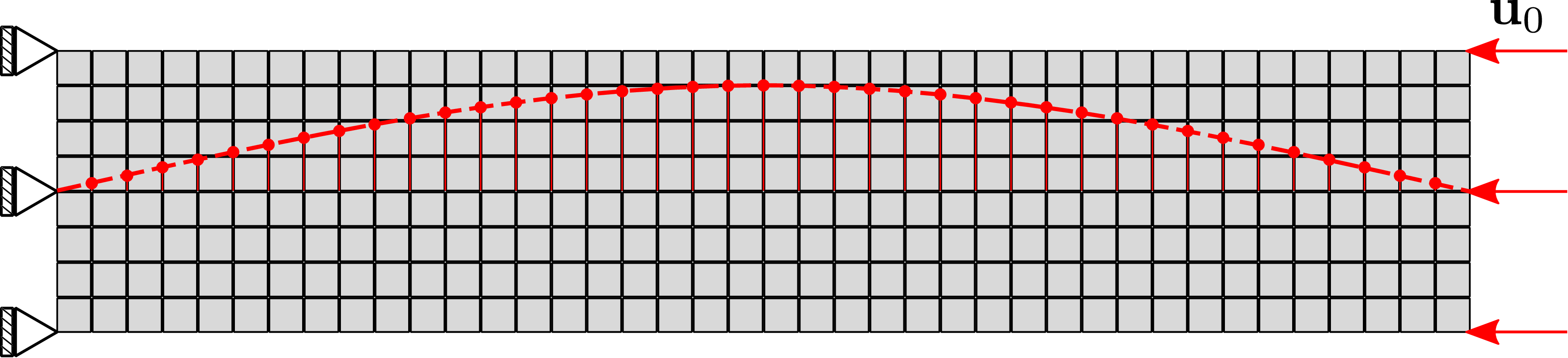}
        \caption{Target bump profile ($40 \times 8$)}
        \label{fig:ex_Umouth_bcs}
    \end{subfigure}
    \caption{Problem settings of the shape matching examples.}
    \label{fig:ex_target_bcs}
\end{figure}

Heterogeneous structures show great potential for applications where a specific deformation pattern is desired upon actuation, e.g., in form-fitting wearables and soft robots~\citep{Mirzaali2018softdevice,Lumpe2021reversemorph,Boley2019shapeshift}.
Motivated by these applications, we optimize two shape matching structures: (1) the target sine-wave deformation profile shown in Fig.~\ref{fig:ex_Uysine_bcs}, and (2) the bump profile in Fig.~\ref{fig:ex_Umouth_bcs}. The first is a cantilever discretized into $30 \times 4$ microstructures, fixed at its left side and loaded with displacement boundary conditions on the right~--~the same example we tested in~\citep{Chan2020metaset}. The second is similar, but discretized into $40 \times 8$. Like the compliance examples, we will also use the truss and freeform basis classes with $M=\{2,3\}$.

In past works~\citep{Chan2020metaset,Wang2020cmame}, we found that these target displacement problems are similar to compliant mechanism design and most effectively solved via a two-stage "top-down" approach that first utilizes inverse TO to find the target effective properties for each microstructure. Departing from previous works, we use the proposed multiclass blending framework in the second stage to optimize the new classes, $\mathbf{c}^{(m)}$, their distributions, $\boldsymbol{\xi}^{(p)}$, and the volume fractions, $\mathbf{v}$, until the target properties are matched.

To find the target properties, i.e., the effective stiffness matrices $\mathbf{C}^H_{t}$, that achieve a desired displacement profile, the first stage follows the method in \cite{Wang2020smo} with following problem:
\begin{equation}\label{eq:to_props}
\begin{aligned}
    & \underset{\mathbf{C}^H_{t}}{\text{minimize}}
    & & \frac{1}{n} \lVert \mathbf{u}-\mathbf{u}_t \rVert _2^2, \\
    & \text{subject to}
    & & \mathbf{K}\mathbf{U} = \mathbf{F},\\
    & & & -\phi(\mathbf{C}^H_{t}) \leq 0,
\end{aligned}
\end{equation}
where $\mathbf{u}$ is the displacement vector of $n$ nodes located on the horizontal centerline of the structure, $\mathbf{u}_t$ is the vector of target displacements of the same nodes, $\mathbf{K}$ is the global stiffness matrix, and $\mathbf{U}$ and $\mathbf{F}$ are global displacement and loading vectors, respectively. 
This inverse problem uses the stiffness matrices of each microstructure as design variables. To ensure that these are within the bounds attainable by shape blending, they are constrained by the signed $L_2$ distance field $\phi$ of the properties of the training data for the neural networks (Sec.~\ref{sec:method_nn} and Fig.~\ref{fig:blend_propspace}c and~f). 

After this, the multiclass FGS is optimized to meet the effective property targets by leveraging our proposed blending scheme. Since we do not aim for a target volume here, there is no global volume constraint and the macrostructure defined by $\mathbf{x}$ remains fixed. The second stage is thus:
\begin{equation}\label{eq:to_match_props}
\begin{aligned}
    & \underset{\mathbf{c},\mathbf{v},\boldsymbol{\xi}}{\text{minimize}}
    & & \frac{1}{N_{el}} \lVert \mathbf{C}^H(\mathbf{c},\mathbf{v},\boldsymbol{\xi})-\mathbf{C}^H_t \rVert _2^2 + kf_{div}(\mathbf{c}), \\
    & \text{subject to}
    & & \mathbf{K}\mathbf{U} = \mathbf{F},
\end{aligned}
\end{equation}
where the bounds on the design variables are the same as in the previous examples. Due to the omission of $\mathbf{x}$, it can be solved with just MMA.
The penalization parameter is set such that $kf_{div}=4$ for the 2-class study and $9$ for 3-class, respectively. The filter radius is $r_{min}=1.2$ and $2.5$ for the sine and bump problems. The volume fractions are initialized at $\mathbf{v}=0.5$, and all other parameters are the same as in the compliance examples.

\begin{figure}[b!]
    \centering
    \begin{subfigure}{0.9\columnwidth}
        \includegraphics[width=1\columnwidth]{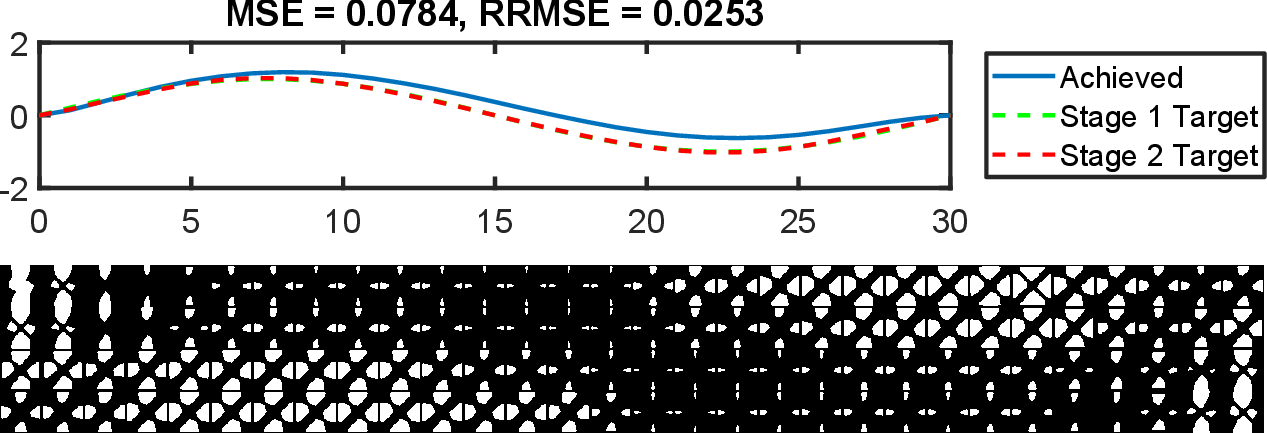}
        \caption{Truss-type, $M=2$}
        \label{fig:ex_target_sine_truss_2}
    \end{subfigure} %
    
    \begin{subfigure}{0.9\columnwidth}
        \includegraphics[width=1\columnwidth]{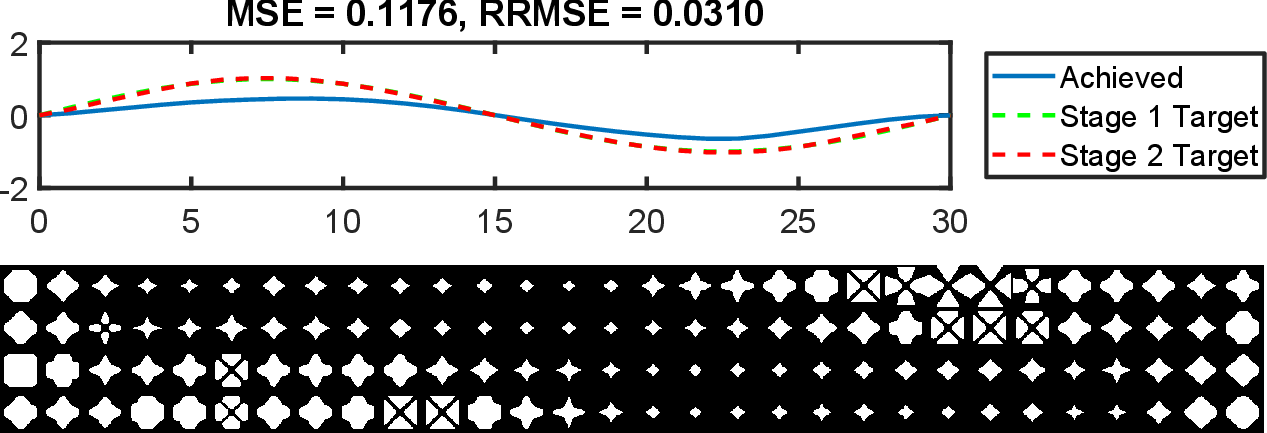}
        \caption{Truss-type, $M=3$}
        \label{fig:ex_target_sine_truss_3}
    \end{subfigure} %
    
    \begin{subfigure}{0.9\columnwidth}
        \includegraphics[width=1\columnwidth]{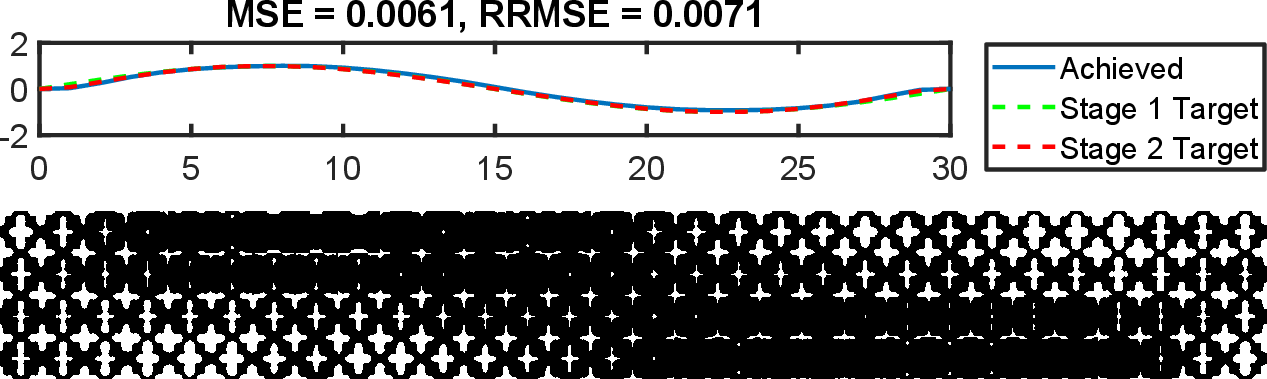}
        \caption{Freeform, $M=2$}
        \label{fig:ex_target_sine_sp20_2}
    \end{subfigure} %
    
    \begin{subfigure}{0.9\columnwidth}
        \includegraphics[width=1\columnwidth]{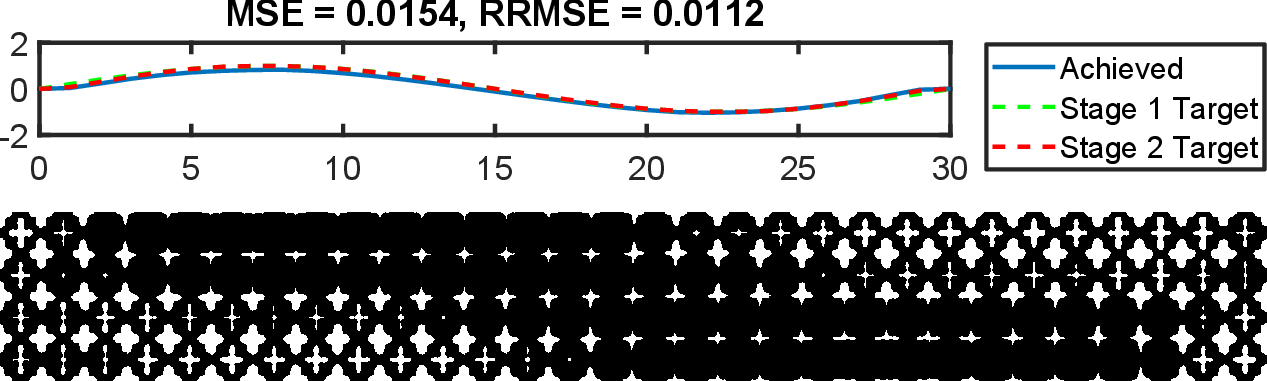}
        \caption{Freeform, $M=3$}
        \label{fig:ex_target_sine_sp20_3}
    \end{subfigure} 
    \caption{Results of the target sine wave problem.}
    \label{fig:ex_target_sine}
\end{figure}

\begin{figure}[th]
    \centering
    \begin{subfigure}{0.9\columnwidth}
        \includegraphics[width=1\columnwidth]{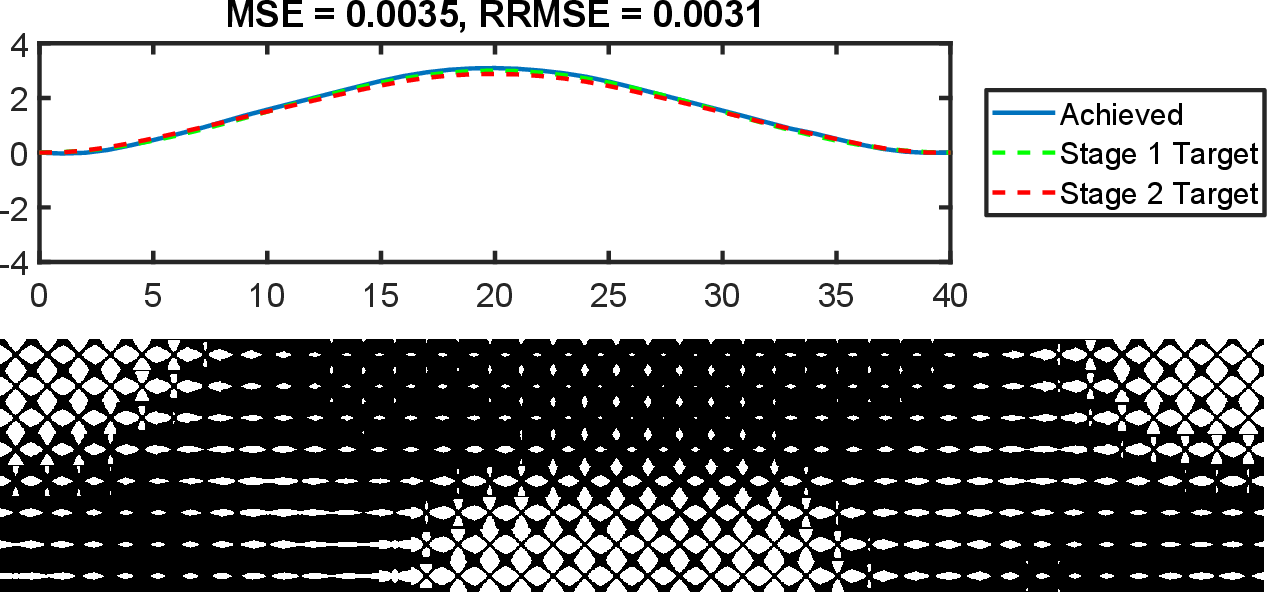}
        \caption{Truss-type, $M=2$}
        \label{fig:ex_target_mouth_truss_2}
    \end{subfigure} %
    
    \begin{subfigure}{0.9\columnwidth}
        \includegraphics[width=1\columnwidth]{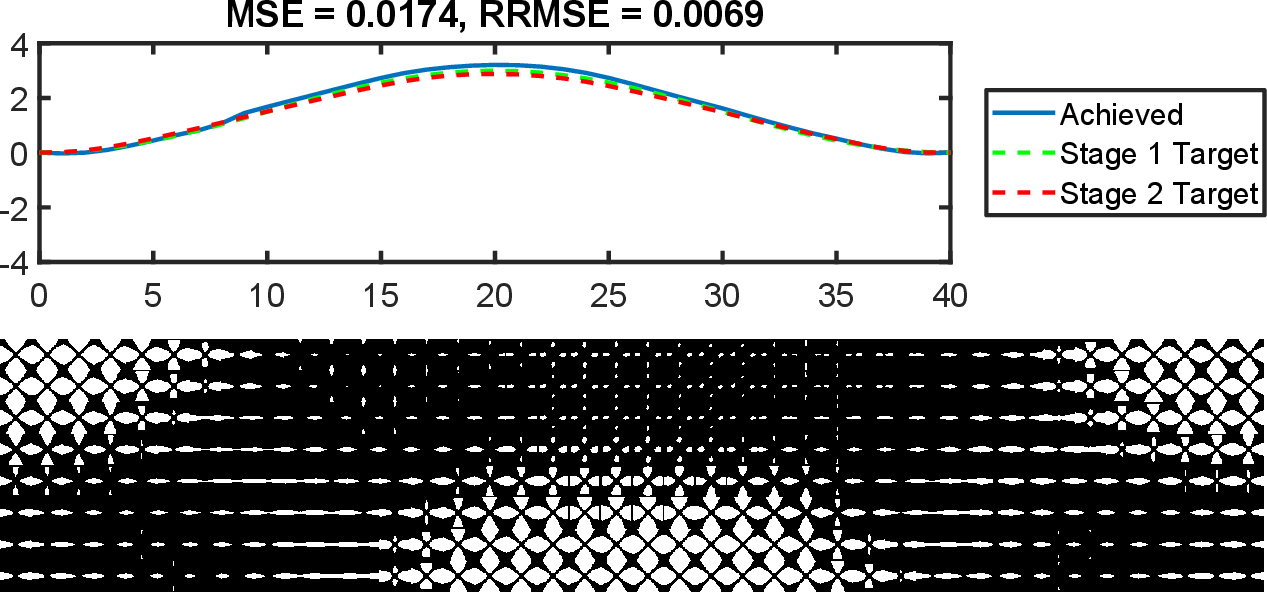}
        \caption{Truss-type, $M=3$}
        \label{fig:ex_target_mouth_truss_3}
    \end{subfigure} %
    
    \begin{subfigure}{0.9\columnwidth}
        \includegraphics[width=1\columnwidth]{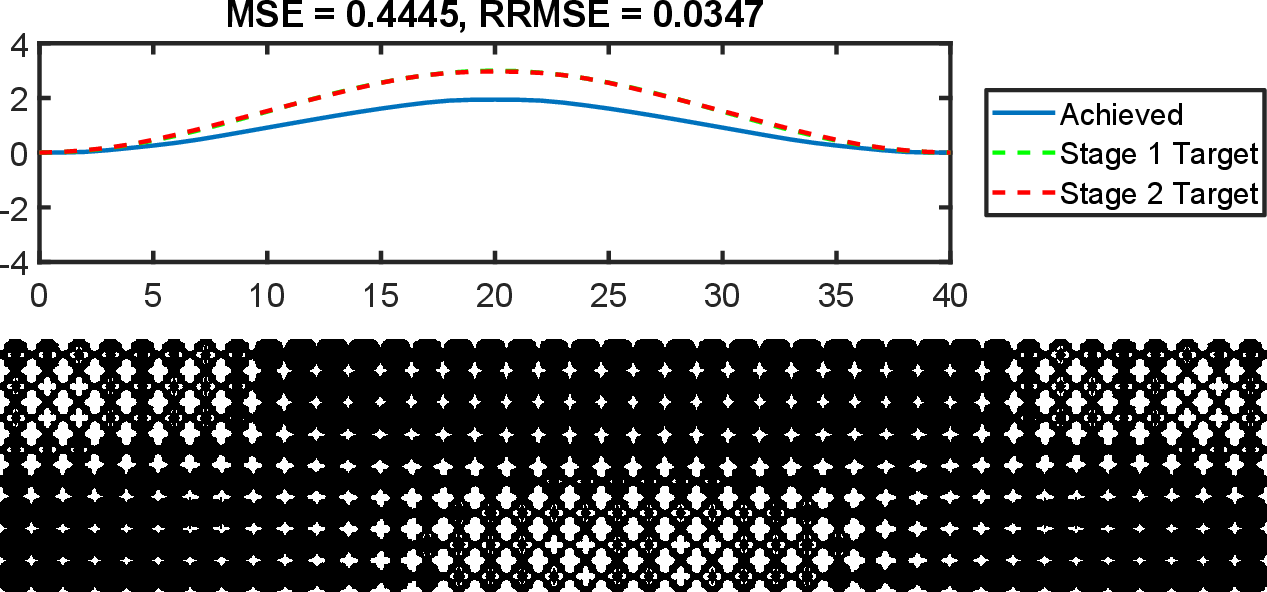}
        \caption{Freeform, $M=2$}
        \label{fig:ex_target_mouth_sp20_2}
    \end{subfigure} %
    
    \begin{subfigure}{0.9\columnwidth}
        \includegraphics[width=1\columnwidth]{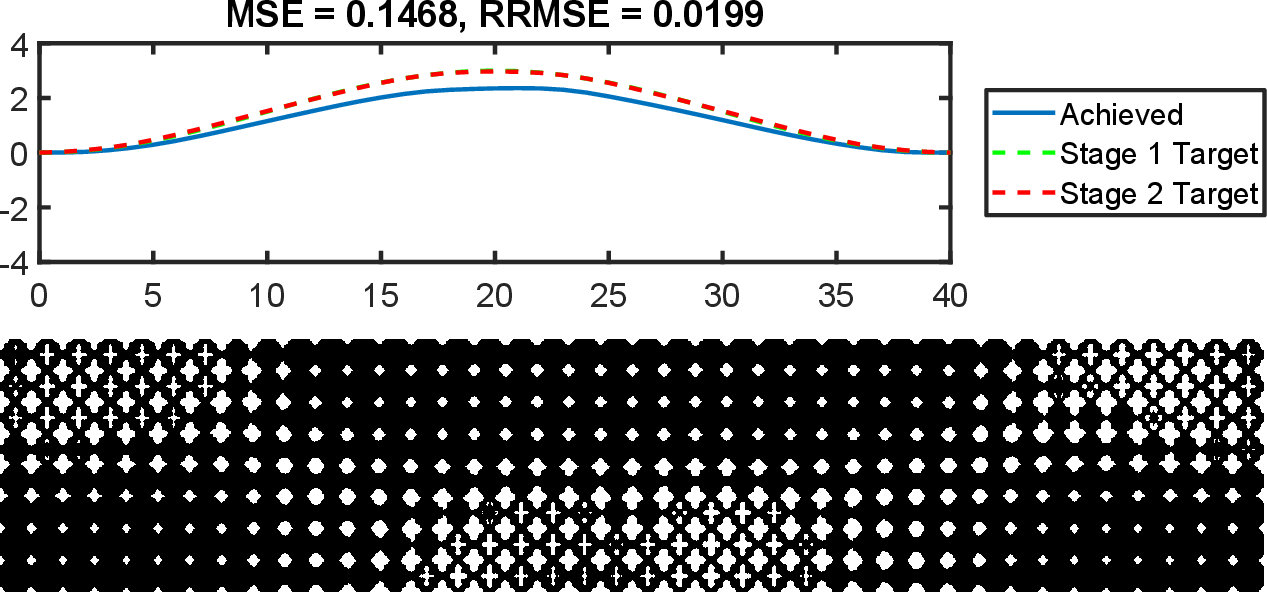}
        \caption{Freeform, $M=3$}
        \label{fig:ex_target_mouth_sp20_3}
    \end{subfigure}
    \caption{Results of the target bump shape example.}
    \label{fig:ex_target_mouth}
\end{figure}

The final multiclass FGS and their achieved displacement profiles (solid blue) are gathered in Fig.~\ref{fig:ex_target_sine} for the sine wave and Fig.~\ref{fig:ex_target_mouth} for the bump problem. In the plots of the displacements, we also show the initial target profile, $\mathbf{u}_t$, used in stage one in dashed green lines, along with profile realized by the optimized properties (dashed red), which serves as an indirect target profile in the second stage.

We performed the same sine wave study in \cite{Chan2020metaset}, but with a combinatorial method for aperiodic designs, i.e., without functional grading. There, the lowest MSE that we achieved was $0.1146$, which most of our proposed multiclass FGS surpass. Interestingly, the freeform basis classes perform considerably better than the truss-types. The freeform designs are composed mostly of the first basis, which we noted in Sec.~\ref{sec:ex_models} has one of the greatest effective Poisson's ratios, as well as the fifth freeform basis, which has both high stiffness and medium Poisson's ratio.

Conversely, the truss-type FGS match the target bump profile more closely than the freeform ones by utilizing the fourth (horizontal) truss basis. 
By inspecting the target properties for each problem, we find that the sine wave requires middling values of both Poisson's ratio and stiffness throughout the FGS, which the freeform classes provide more easily, whereas the bump profile needs distinct regions of either large $x$-directional stiffness or high Poisson's ratio, which the first and fourth truss classes meet exceptionally well (see Fig.~\ref{fig:blend_propspace}c and~f). 
This observation portends a possible extension of our work where the most efficient basis classes can be chosen to match the initial distribution of principal macroscale stresses for specific problems, similar to \cite{Xu2018clusterstress}.

Another intriguing note is that, by blending the last two truss basis classes, we can form square microstructures that are not found in the original set (Fig.~\ref{fig:ex_target_sine_truss_3}). Moreover, combining those two with the 'X' basis creates microstructures with star-shaped voids that are not strictly trusses. These are direct results of the weighted sum of SDFs in our proposed blending scheme (Eq.~\ref{eq:blend_inner}), which can non-intuitively morph the basis classes to achieve optimal performance.

\vfill
\section{Conclusions}\label{sec:conclusion}
We proposed in this work a novel multiclass shape blending scheme that provides a low-dimensional representation of microstructures for both design and DL, and a data-driven multiscale design method that utilizes a hybrid of TO algorithms along with a new penalty on low diversity designs. By integrating these, we created a multiclass FGS design framework that encapsulates the freedom of fully aperiodic structures while featuring efficiency greater than that of typical multiscale methods. 
The key is the ability of shape blending to blur the lines between classes, creating graded designs with novel microstructures beyond the initial basis classes. Even with classes that have complex features or are not mutually compatible, continuous transitions between neighboring microstructures are guaranteed. 

Furthermore, feasibility constraints are incorporated into the scheme to ensure that they are naturally met. In this work, we use a simple measure~--~minimum feature size. However, defining the lower feasible bounds of each basis class outside of the optimization process means there is potential for future works to incorporate other feasibility, or even quality, metrics, such as those without cheap or tractable gradients. 

We demonstrated these advantages through compliance and shape matching examples, in which blending empowered our FGS to surpass designs in literature. Our results revealed that truss-type classes consistently achieve low compliance, and that diverse freeform classes reach satisfactory performance across multiple applications despite being automatically chosen without considering their compatibility. We also discovered that more is not always better when it comes to classes. By encouraging the design to converge to a smaller number of diverse classes, as few as two can be blended to obtain optimal designs. This outcome merits deeper exploration in the future on how diversity metrics can benefit structural design.

Our framework is general in that it is not tied to the specific DL and TO methods shown in this work. It is also not limited to our 2D classes, since multiclass blending is independent of the topology, representation, dimension and resolution of the basis shapes. This modularity is an especially welcome feature as more advanced prediction models and TO algorithms emerge to solve complex multiphysics and nonlinear mechanics problems, including 3D ones. 
Beyond the examples presented, our framework can be extended to sought-after functionalities like thermoelasticity, fracture resistance and energy absorption, and adapted to applications such as customized user products and architectural design. 
We believe these are all exciting avenues for future works enabled by multiclass shape blending.

\section*{Acknowledgements}
We are grateful for the MMA codes from Prof. K. Svanberg at the Royal Institute of Technology, and the BESO codes from Profs. X. Huang and Y. M. Xie at RMIT University.

\section*{Funding information}
The authors were supported by the National Science Foundation (NSF) CSSI program (Grant No.~OAC-1835782). Yu-Chin Chan received funding from the NSF Graduate Research Fellowship (Grant No.~DGE-1842165), and Liwei Wang from the Zhiyuan Honors Program for Graduate Students of Shanghai Jiao Tong University for his predoctoral visit at Northwestern University.

\section*{Conflict of interest}
The authors declare that they have no conflict of interest.

\section*{Replication of results}
All proposed schemes and algorithms are disclosed within this paper and developed by the authors in MATLAB (multiclass shape blending, concurrent design, data generation, deep learning) and Python (pre-processing of basis classes). Additional codes are adapted from open-source packages and are cited. The dataset from which the diverse freeform basis classes are extracted can be found at \url{https://ideal.mech.northwestern.edu/}.

\vfill\break

\begin{appendices}
\section{Algorithms}\label{sec:appen_algs}

\begin{algorithm}[!h]
\caption{Adaptive scheme to decrease volume fraction limits during concurrent multiscale design. $i$ denotes the number of the current iteraton.}\label{alg:alg_mcfgs_vol} 
\begin{algorithmic}[1]
\Require $V^*_{Global,i-1}, V^*_{BESO,i-1}$, $V_{Global}, V_{BESO}$;
\If{$(i \mod 10) = 0$ \textbf{and} $V^*_{Global,i-1}>V^*_{Global}$}
    \State $V^*_{Global,i} \leftarrow \min(V_{Global},V^*_{Global,i-1})-0.025$;
\EndIf
\If{$(i \mod 10) = 0$ \textbf{and} $V_{BESO}\leq V^*_{BESO,i-1}$}
    \State $V^*_{BESO,i} \leftarrow V^*_{BESO,i-1}-0.005$;
\EndIf
\State \Return updated volume constraints.
\end{algorithmic}
\end{algorithm}

\algnewcommand{\Initialize}[1]{%
  \State \textbf{Initialize:}
  \Statex \hspace*{\algorithmicindent}\parbox[t]{.8\linewidth}{\raggedright #1}
}
\begin{algorithm}[!h]
\caption{Concurrent design framework for multiclass functionally graded structures. If there are no volume constraints, ignore Line 10. If the macrostructure is fixed, ignore Line 12.}\label{alg:alg_mcfgs} 
\begin{algorithmic}[1]
\Initialize{design variables $\mathbf{c},\mathbf{v},\boldsymbol{\xi},\mathbf{x}$; \\
volume constraints $V^*_{Global,0}, V^*_{BESO,0}$; \\
weight on low-diversity penalty $k$; }
\While{change in design $>tol$}
    \State $i\leftarrow i + 1$; \Comment{iteration counter}
    \For{each macro-element $e$}
        \State obtain $\hat{\mathbf{c}}_e$ (Eqs.~\ref{eq:class_interp},~\ref{eq:global_interp});
        \State find $\boldsymbol{\Phi}_e$ and $t_e$ so that unit cell has volume $\hat{v}_e$ (Eq.~\ref{eq:blend_final});
        \State approximate $\hat{v}^a_e$ (Eqs.~\ref{eq:sigmoid},~\ref{eq:sig_vf});
        \State predict effective stiffness $\mathbf{C}^H_e=NN(\bar{\mathbf{c}}_e,\hat{v}_e)$ and obtain $\mathbf{k}_e$;
    \EndFor
   
    \State update volume fraction constraint limits (Algorithm~\ref{alg:alg_mcfgs_vol});
    \State compute objective, constraints and sensitivities (Appendix~\ref{sec:appen_sens});
    \State update macroscale design $\mathbf{x}$ with BESO;
    \State update other variables $\mathbf{c},\mathbf{v},\boldsymbol{\xi}$ with MMA;
\EndWhile
\State \Return optimal multiclass functionally graded design.
\end{algorithmic}
\end{algorithm}

\renewcommand{\thefigure}{\arabic{figure}}
\setcounter{figure}{14}
\renewcommand{\thetable}{\arabic{table}}
\setcounter{table}{4}
\newpage
\section{Additional Results}\label{sec:appen_results}

\begin{figure*}[th!]
    \centering
    \includegraphics[width=0.85\textwidth]{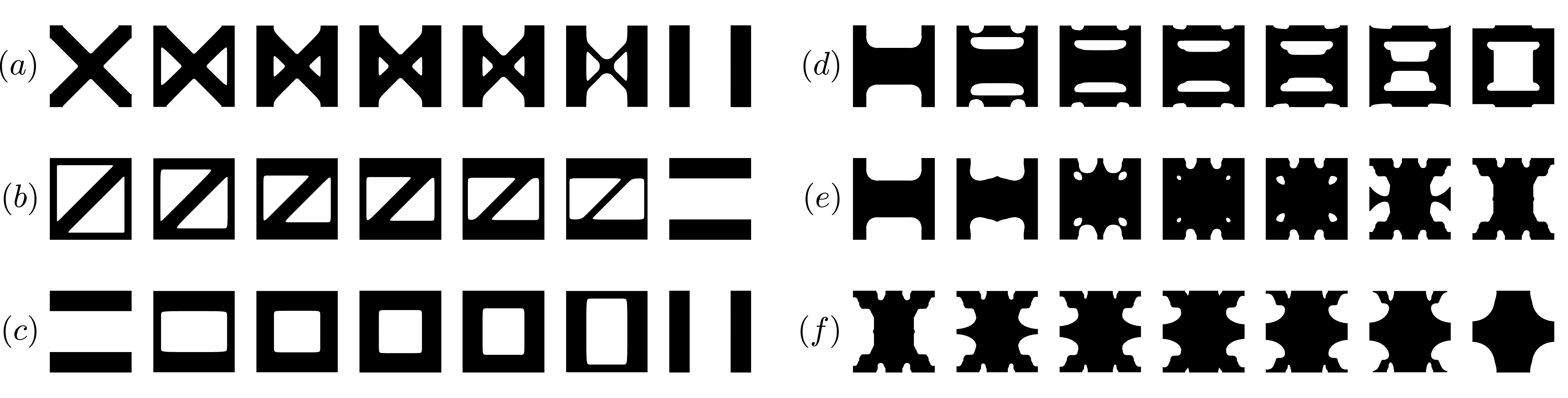}
    \caption{Examples of blending pairs of basis classes. The first and last microstructures are from basis classes, and those in between are produced by linearly interpolating the weights in the blending scheme. Between truss classes (left column): (a) 1 and 5, (b) 2 and 4, and (c) 4 and 5. Between freeform classes (right column): (d) 2 and 3, (e) 2 and 4, and (f) 4 and 5.}
    \label{fig:blend_demo_selected}
\end{figure*}

\subsection{Further Examples of Blending}
To help visualize blending between pairs of basis classes, we illustrate several examples in Fig.~\ref{fig:blend_demo_selected}. In particular, we select pairs that have not appeared in our design results. No matter how simple, complex or incompatible the basis classes, our multiclass blending scheme is able to provide feasible, i.e., well-connected, intermediate microstructures. This is the benefit of (1) the interpolation between the SDFs of basis classes, and (2) the imposition of the lower feasible bounds of each basis through Eqs.~\ref{eq:blend_inner} and~\ref{eq:blend_outer}.

\subsection{Compliance Validation}
In Sec.~\ref{sec:ex_mbb}, we presented the compliance as calculated by numerical homogenization to directly compare our results to those in literature. In Tables~\ref{tab:comp_compare_mbb} and~\ref{tab:comp_compare_bridge}, we additionally report the compliance of our designs performed using: (1) neural networks (same as during design), (2) numerical homogenization (as shown in the main paper), and (3) fine mesh analysis using the multigrid method~\citep{Amir2013mgcg} (for validation).

For MBB in particular, the homogenization-based compliance agree reasonably well with the fine mesh analysis. In general, the numerical homogenization-based results, $f_{c,Hom}$, are slightly closer to the fine analysis values, $f_{c,Fine}$, than the ones using the neural networks, $f_{c,NN}$, which is expected. However, the differences between $f_{c,NN}$ and $f_{c,Hom}$ are not large, validating the use of our predictive models to accelerate the design process.

\begin{table}[th]
\centering
\caption{Comparison of the compliance of our MBB results in Sec.~\ref{sec:ex_mbb_M2} calculated by: neural network models $f_{c,NN}$, numerical homogenization $f_{c,Hom}$, fine mesh analysis $f_{c,Fine}$. For the homogenization-based values, we also report the percent error from $f_{c,Fine}$ in parentheses.}
\label{tab:comp_compare_mbb}
\begin{tabular}{@{}cclll@{}}
\toprule
Basis                     & M & $f_{c,NN}$ & $f_{c,Hom}$ & $f_{c,Fine}$ \\ \midrule
\multirow{2}{*}{Truss}    & 2 & \begin{tabular}[c]{@{}l@{}}$187.41$\\ (-$13.44\%$)\end{tabular} & \begin{tabular}[c]{@{}l@{}}$192.47$\\ (-$11.10\%$)\end{tabular} & $216.50$ \\ \cmidrule(l){2-5} 
                          & 3 & \begin{tabular}[c]{@{}l@{}}$199.74$\\ (-$4.56\%$)\end{tabular} & \begin{tabular}[c]{@{}l@{}}$188.85$\\ (-$ 9.76\%$)\end{tabular} & $209.28$           \\ \midrule
\multirow{2}{*}{Freeform} & 2 & \begin{tabular}[c]{@{}l@{}}$210.52$\\ (-$ 20.64\%$)\end{tabular} & \begin{tabular}[c]{@{}l@{}}$223.13$\\ (-$ 15.89\%$)\end{tabular} & $265.28$           \\ \cmidrule(l){2-5} 
                          & 3 & \begin{tabular}[c]{@{}l@{}}$212.68$\\ (-$ 22.55\%$)\end{tabular} & \begin{tabular}[c]{@{}l@{}}$229.15$\\ (-$ 16.55\%$)\end{tabular} & $274.60$           \\ \bottomrule 
\end{tabular}
\end{table}

\begin{table}[th!]
\centering
\caption{Compliance of our bridge results in Sec.~\ref{sec:ex_mbb_M3} calculated three different ways, including the percent error from $f_{c,Fine}$ in parentheses.}
\label{tab:comp_compare_bridge}
\begin{tabular}{@{}cclll@{}}
\toprule
Basis                     & M & $f_{c,NN}$ & $f_{c,Hom}$ & $f_{c,Fine}$ \\ \midrule
\multirow{2}{*}{Truss}    & 2 & \begin{tabular}[c]{@{}l@{}}$90.00$\\ (-$ 34.92\%$)\end{tabular} & \begin{tabular}[c]{@{}l@{}}$92.22$\\ (-$ 33.31\%$)\end{tabular} & $138.29$           \\ \cmidrule(l){2-5} 
                          & 3 & \begin{tabular}[c]{@{}l@{}}$92.59$\\ (-$ 31.70\%$)\end{tabular} & \begin{tabular}[c]{@{}l@{}}$94.74$\\ (-$ 30.12\%$)\end{tabular} & $135.57$           \\ \midrule
\multirow{2}{*}{Freeform} & 2 & \begin{tabular}[c]{@{}l@{}}$95.65$\\ (-$ 33.58\%$)\end{tabular} & \begin{tabular}[c]{@{}l@{}}$100.15$\\ (-$ 30.46\%$)\end{tabular} & $144.01$           \\ \cmidrule(l){2-5} 
                          & 3 & \begin{tabular}[c]{@{}l@{}}$93.12$\\ (-$ 36.51\%$)\end{tabular} & \begin{tabular}[c]{@{}l@{}}$102.24$\\ (-$ 30.29\%$)\end{tabular} & $146.66$           \\ \bottomrule 
\end{tabular}
\end{table}

\begin{figure*}[th!]
    \centering
    \includegraphics[width=0.97\textwidth]{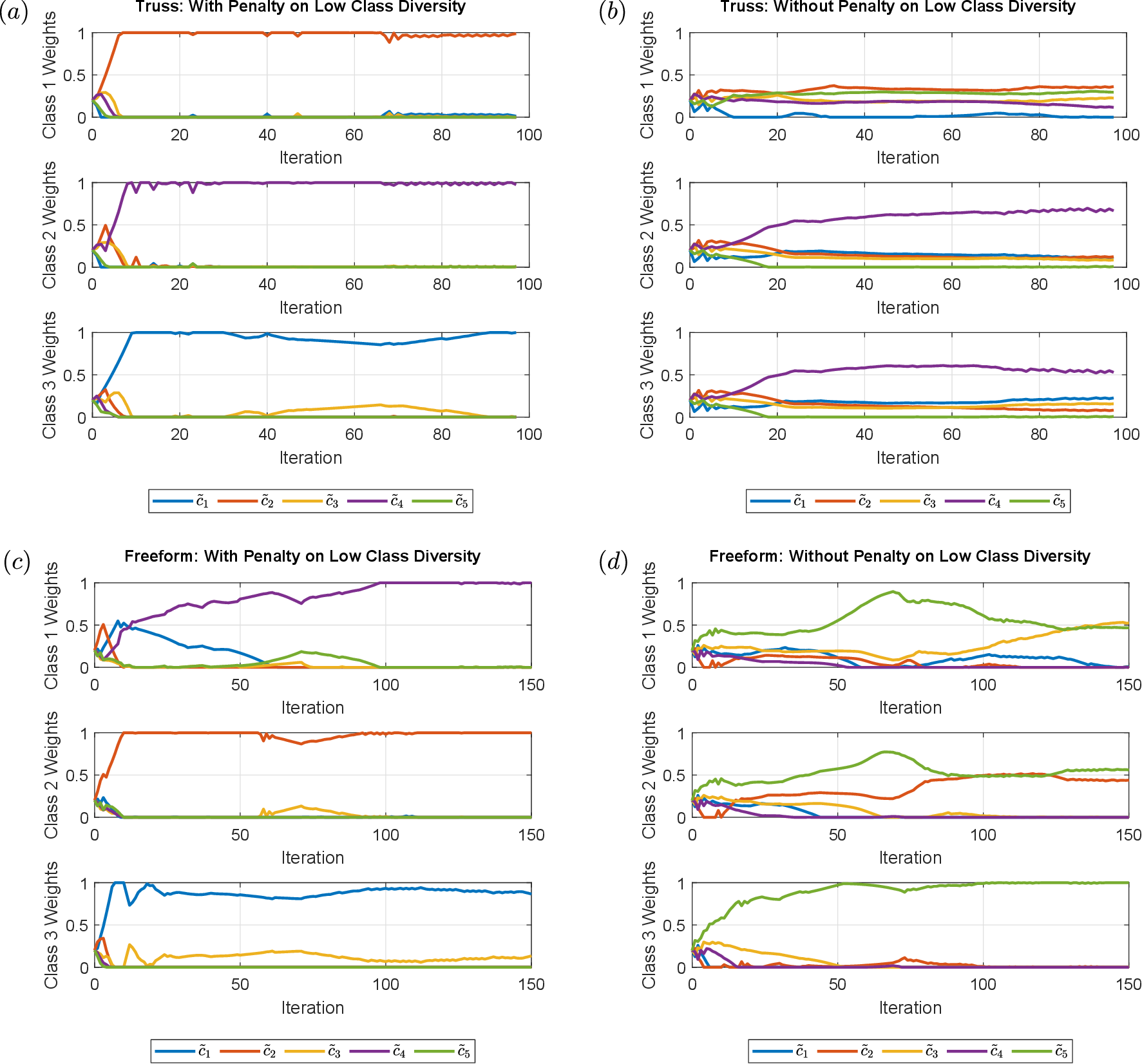}
    \caption{For the 3-class MBB example, convergence plots of the class design variables with the proposed penalty (left column) and without (right column).}
    \label{fig:conv_classes}
\end{figure*}

\subsection{Comparison of Convergence With and Without Penalty}
To demonstrate the effect of the low-diversity penalty (Sec.~\ref{sec:method_design_div}) on convergence, we compare the 3-class MBB examples using both the freeform and truss basis classes (Sec.~\ref{sec:ex_mbb_M3}). This problem had the most difficulty converging to low compliance values. The histories of the class design variables (which are the only variables used to compute the penalty) are plotted in Fig.~\ref{fig:conv_classes}.

The penalty on low class diversity has a significant impact on the convergence behavior of the class design variables. They confirm our statement in Sec.~\ref{sec:ex_mbb_div} that applying the penalty can help the weights of the class design variables to converge more quickly, as well as alleviate possible local minima (e.g., compare Figs.~\ref{fig:conv_classes} a and b).

\section{Sensitivity Analysis}\label{sec:appen_sens}
The derivations of the sensitivity analysis for gradient-based topology optimization (TO) are detailed in this section. The nomenclature are the same as in the main text. All derivations were verified using the finite difference method.

For the global multiclass shape blending scheme (Eq.~\ref{eq:blend_final}), which is used to obtain the microscale topology at $e$, the gradient is
\begin{equation}\label{eq:blend_dchat}
\begin{aligned}
    \frac{\partial \boldsymbol{\Phi}_e}{\partial \hat{c}_{e \cdot d}} =     &
    \frac{1}{\beta_2 \boldsymbol{\Phi}_d^I}
    \Big[ \beta_2 \exp{\big( \beta_2 \boldsymbol{\Phi}_e^0 \big)} \odot \boldsymbol{\Phi}_d^{*} \\     &
    + \frac{\partial H(\hat{c}_{e \cdot d},\beta_2,\eta_2)}{\partial \hat{c}_{e \cdot d}} \exp{\big( \beta_2 \boldsymbol{\Phi}_d^{L} \big)} \Big],
\end{aligned}
\end{equation}
where
\begin{equation}\label{eq:blend_dchat_inner}
    \boldsymbol{\Phi}_e^I = 
    \exp{\big( \beta_2 \boldsymbol{\Phi}_e^0 \big)} 
    + \sum_d^D a_d \exp{\big( \beta_2 \boldsymbol{\Phi}_d^{L} \big)},
\end{equation}
\begin{equation}\label{eq:blend_inner_final}
    \boldsymbol{\Phi}_e^0 = \sum_d^D \hat{c}_{e \cdot d} \boldsymbol{\Phi}_d^{*} + t_e,
\end{equation}
\begin{equation}\label{eq:heav_dx}
    \frac{\partial H(\hat{c}_{e \cdot d})}{\partial \hat{c}_{e \cdot d}} = \frac{ \beta_2 \big[ 1 - \tanh{^2 (\beta_2(\hat{c}_{e \cdot d}-\eta_2))} \big] }{ \tanh{(\beta_2\eta_2)} + \tanh{(\beta_2(1-\eta_2))} },
\end{equation}
and $\odot$ indicates element-wise multiplication.

The sensitivities for the the class interpolation schemes (Eqns.~\ref{eq:class_interp} and~\ref{eq:global_interp}) with respect to the class design variables is as follows:
\begin{equation}\label{eq:global_interp_dc}
\begin{aligned}
    & \frac{\partial \hat{c}_{e \cdot d}}{\partial c_d^{(m)}}     
    = \Bigg\{ \frac{\partial \tilde{c}_d^{(1)}}{\partial \Tilde{c}_d^{(m)}} \\    &
    + \sum^{M-1}_{j=1} \bigg[ \frac{\partial \big( \tilde{c}_d^{(j+1)}-\tilde{c}_d^{(j)} \big)}{\partial \Tilde{c}_d^{(m)}} \prod^{j}_{k=1} \hat{\xi}^{(k)}_e \bigg] \Bigg\} \frac{\partial \Tilde{c}_d^{(m)}}{\partial c_d^{(m)}},
\end{aligned}
\end{equation}
where
\begin{equation}\label{eq:class_interp_dc}
    \frac{\partial \Tilde{c}_d^{(m)}}{\partial c_d^{(m)}} = \sum^{D-1}_{j=d} \bigg[ \big(z_d^{(j+1)} - z_d^{(j)}\big) \prod^{j}_{k=1,k \neq d} c_k^{(m)} \bigg].
\end{equation}
With respect to the macroscale distribution fields, it is:
\begin{equation}\label{eq:global_interp_dxi}
    \frac{\partial \hat{c}_{e \cdot d}}{\partial \xi^{(p)}_e} = \sum^{M-1}_{j=p} \bigg[ \big(\tilde{c}_d^{(j+1)} - \tilde{c}_d^{(j)}\big) \prod^{j}_{k=1,k \neq p} \hat{\xi}^{(k)}_e \bigg] \frac{\hat{\xi}^{(p)}_e}{\xi^{(p)}_e}.
\end{equation}

We also need the sensitivities of the radial filters. This follows the density filters of traditional TO methods closely~\citep{sigmund2007filters}, and so we do not repeat it here for brevity.

For compliance (Eq.~\ref{eq:to_compliance}), the adjoint method~\citep{Bendse2004simp} and chain rule allow us to derive the following with respect to the design variables:
\begin{equation} 
    \frac{\partial f_c}{\partial c_{e \cdot d}} = - \Bigg[ \sum_{n-1}^{N_{el}} \mathbf{u}_n^T \frac{\partial \mathbf{k}_n}{\partial \hat{c}_{e \cdot d}} \mathbf{u}_n \Bigg] \frac{\partial \hat{c}_{e \cdot d}}{\partial c_{e \cdot d}},
\end{equation}
\begin{equation} 
    \frac{\partial f_c}{\partial v_e} = - \Bigg[ \sum_{n-1}^{N_{el}} \mathbf{u}_n^T \frac{\partial \mathbf{k}_n}{\partial \hat{v}_e} \mathbf{u}_n \Bigg] \frac{\partial \hat{v}_e}{\partial v_e},
\end{equation}
and
\begin{equation} 
    \frac{\partial f_c}{\partial \xi^{(p)}_e} = - \Bigg[ 
    \sum_{n-1}^{N_{el}} \mathbf{u}_n^T \frac{\partial \mathbf{k}_n}{\partial \hat{\xi}^{(p)}_e} \mathbf{u}_n 
    \Bigg] 
    \frac{\hat{\xi}^{(p)}_e}{\xi^{(p)}_e}.
\end{equation}

The derivatives of the element effective stiffness matrices are
\begin{equation} 
    \frac{\partial \mathbf{k}_e}{\partial \hat{c}_{e \cdot d}} = \frac{ \partial \mathbf{k}_e }{ \partial \mathbf{C}_e } \frac{\partial \mathbf{C}_e}{\partial \hat{c}_{e \cdot d}},
\end{equation} 
\begin{equation} 
    \frac{\partial \mathbf{k}_e}{\partial \hat{v}_e} = \frac{ \partial \mathbf{k}_e }{ \partial \mathbf{C}_e } \frac{\partial \mathbf{C}_e}{\partial \hat{v}_e},
\end{equation}
and
\begin{equation} 
\begin{aligned}
    \frac{\partial \mathbf{k}_e}{\partial \hat{\xi}^{(p)}_e} = \frac{ \partial \mathbf{k}_e }{ \partial \mathbf{C}_e } 
    \Bigg[ 
    \sum_{d=1}^D 
    & \frac{\partial \mathbf{C}_e}{\partial \hat{c}_{e \cdot d}} \frac{\partial \hat{c}_{e \cdot d}}{\partial \hat{\xi}^{(p)}_e} 
    \Bigg].
\end{aligned}
\end{equation} 
Here, due to our data-driven framework, the gradients of the effective properties $\partial \mathbf{C}_e / \partial \hat{c}_{e \cdot d}$ and $\partial \mathbf{C}_e / \partial \hat{v}_e$ are obtained by backpropagating through the layers of the fully connected neural network~\citep{Hastie2009learning}. 

To obtain the sensitivities of the global volume fraction constraint, we can use Eq.~\ref{eq:blend_dchat} and the Sigmoid function (Eq.~\ref{eq:sigmoid} in Sec.~\ref{sec:method_design_vf}), which gives us a continuous approximation of a volume fraction ($\hat{v}^a_e$ from Eq.~\ref{eq:sig_vf}). First, the derivative of the Sigmoid applied to a microstructure's SDF ($\boldsymbol{\Phi}_e$) is
\begin{equation}\label{eq:sigmoid_dphi}
    \frac{\partial S(\boldsymbol{\Phi}_e,\beta_1)}{\partial \boldsymbol{\Phi}_e}  = \beta_1 S(\boldsymbol{\Phi}_e,\beta_1) \big( 1-S(\boldsymbol{\Phi}_e,\beta_1) \big).
\end{equation}

Therefore, the sensitivity of the global volume fraction can be decomposed as
\begin{equation}
\begin{aligned}
    \frac{\partial V_{Global}}{\partial \hat{c}_{e \cdot d}} &= \frac{\partial V_{Global}}{\partial \hat{v}^a_e} \frac{\partial \hat{v}^a_e}{\partial S} \frac{\partial S}{\partial \boldsymbol{\Phi}_e} \frac{\partial \boldsymbol{\Phi}_e}{\partial \hat{c}_{e \cdot d}}, \\
    &= \frac{x_e}{N_{el}} \bigg(\frac{1}{n_{el}} \sum_{u=1}^{n_{el}} \frac{\partial S}{\partial \boldsymbol{\Phi}_e} \bigg) \frac{\partial \boldsymbol{\Phi}_e}{\partial \hat{c}_{e \cdot d}},
\end{aligned}
\end{equation}
and the sensitivities of the constraint itself with respect to the design variables are:
\begin{equation}
    \frac{\partial g_1}{\partial c_{e \cdot d}} = \frac{1}{V^*_{Global}}  \frac{\partial V_{Global}}{\partial \hat{c}_{e \cdot d}} \frac{\partial \hat{c}_{e \cdot d}}{\partial c_{e \cdot d}},
\end{equation}
\begin{equation}
    \frac{\partial g_1}{\partial v_e} = \frac{1}{V^*_{Global}},
\end{equation}
and
\begin{equation}
    \frac{\partial g_1}{\partial \xi^{(p)}_e} = \frac{1}{V^*_{Global}} \frac{\partial V_{Global}}{\partial \hat{c}_{e \cdot d}} \frac{\partial \hat{c}_{e \cdot d}}{\xi^{(p)}_e}.
\end{equation}

We note that these approximations of the microstructural volume fractions do introduce some error into the sensitivities but, in our experience, are minor.

Since we use the default algorithm for BESO developed by \cite{Huang2007beso} to update $\mathbf{x}$, the filtering of the macroscale sensitivity numbers for the compliance problems can be found in the original paper. The only difference is that we use the effective stiffness predicted by the data-driven models. Thus, our sensitivity numbers are modified to be
\begin{equation}
    \alpha_e = \mathbf{u}_e^T \mathbf{k}^H_e(\hat{\mathbf{c}}_e,\hat{v}_e) \mathbf{u}_e.
\end{equation}

For the low-diversity penalty function, the sensitivity can be derived using chain rule by taking the gradient of the log-determinant~\citep{boyd2004convex}, then the Gaussian kernel and Euclidean distance between the class design variables, $\mathbf{c}^{(m)}$.

Finally, for the derivations of the first stage of the shape matching problem (Eq.~\ref{eq:to_props}), we refer the reader to our previous work~\citep{Wang2020cmame}. For the second stage (Eq.~\ref{eq:to_match_props}) the derivations are straightforward to calculate by following the same steps above, substituting compliance for the mean squared error (MSE) between target and designed effective properties.

\end{appendices}

\vfill\break
\bibliography{ms}

\end{document}